\documentclass[aps,showpacs,showkeys,eqsecnum]{revtex4}
\usepackage{amsfonts,amsmath,amssymb,bm,graphicx}

\usepackage{slashed} 
\usepackage{anysize} 
\marginsize{2cm}{2cm}{1cm}{1cm} 
\usepackage{bm} 
\usepackage{amsfonts} 
\usepackage{amsmath} 
\usepackage{amssymb} 
\usepackage{graphicx} 
\usepackage[usenames,dvipsnames]{xcolor} 
\usepackage[colorlinks=true,linkcolor=Blue,citecolor=Blue,linktocpage,urlcolor=Blue]{hyperref} 
\usepackage{epstopdf} 
 
\newcommand{\s}{\mathbf{\sigma}}
\newcommand{\m}{\mathbf{\mu}}
 
\newcommand{\aeq}{\begin{equation}} 
\newcommand{\ceq}{\end{equation}} 
\newcommand{\aec}{\begin{eqnarray}} 
\newcommand{\cec}{\end{eqnarray}} 
\newcommand{\ase}{\begin{subequations}} 
\newcommand{\cse}{\end{subequations}} 

\renewcommand{\(}{\left(} 
\renewcommand{\)}{\right)} 
\renewcommand{\[}{\left[} 
\renewcommand{\]}{\right]}

\renewcommand{\a}{\alpha} 
\renewcommand{\b}{\beta} 

\newcommand{\n}{\nu} 
\renewcommand{\o}{\omega} 
\newcommand{\g}{\gamma} 
\renewcommand{\d}{\delta} 
\newcommand{\h}{\eta} 
\newcommand{\z}{\zeta}

\newcommand{\y}{\psi} 
\renewcommand{\l}{\lambda} 

\newcommand{\q}{\theta} 
\newcommand{\e}{\epsilon} 
\renewcommand{\t}{\tau} 
\newcommand{\p}{\pi} 

\newcommand{\G}{\Gamma}

\newcommand{\D}{\Delta}

\newcommand{\pd}{\partial} 
\newcommand{\textlineskip}{\baselineskip=13pt}
\newcommand{\smalllineskip}{\baselineskip=10pt}

\DeclareFontFamily{OT1}{pzc}{} 
\DeclareFontShape{OT1}{pzc}{m}{it}{<-> s * [1.10] pzcmi7t}{} 
\DeclareMathAlphabet{\mathpzc}{OT1}{pzc}{m}{it} 
\oddsidemargin=\evensidemargin
\textlineskip
\smalllineskip

\begin{document}



\title{Gyromagnetic $g_s$ factors of the spin-$1/2$ particles  in the
 ${\Big(}1/2^+$-$1/2^-$-$3/2^-{\Big)}$ triad of the 
four-vector spinor,  $\psi_\mu$, irreducibility,  and linearity}
\author{ E. G. Delgado Acosta}\email{german@ifisica.uaslp.mx}

\affiliation{Institute of Physics, Autonomous University at San Luis Potosi,\\
Av. Manuel Nava 6, University Campus, San Luis Potosi, SLP 78290, Mexico\\
}

\author{V. M. Banda Guzm\'an }\email{vmbg@ifisica. uaslp.mx}

\affiliation{Institute of Physics, Autonomous University at San Luis Potosi,\\
Av. Manuel Nava 6, University Campus, San Luis Potosi, SLP 78290, Mexico\\
}

\author{M. Kirchbach}\email{mariana@ifisica.uaslp.mx}

\affiliation{Institute of Physics, Autonomous University at San Luis Potosi,\\
Av. Manuel Nava 6, University Campus, San Luis Potosi, SLP 78290, Mexico\\
}


\begin{abstract} 
The gauged Klein-Gordon equation, extended by a $g_s \sigma_{\mu\nu}F^{\mu\nu}/4$ interaction,  the contraction of the electromagnetic field strength tensor,
$F^{\mu\nu}$, with the generators, $\sigma_{\mu\nu}/2$, of the Lorentz group in  $(1/2,0)\oplus (0,1/2)$, and  $g_s$ 
being the  gyromagentic factor, is examined with the aim to find out as to what extent it qualifies as a wave equation for general relativistic spin-$1/2$ particles transforming as $(1/2,0)\oplus (0,1/2)$ and possibly distinct from the Dirac fermions. 
This equation can be viewed as the generalization of the $g_s=2$ case, known under the name of the  Feynman-Gell--Mann equation, 
the only one which allows for a bi-linearization into the gauged Dirac equation and its conjugate. At the same time,  it is well known a fact that a $g_s=2$ value can also be obtained  upon the bi-linearization of the non-relativistic Schr\"odinger into non-relativistic  Pauli equations. The inevitable conclusion is that it must not be necessarily  relativity which fixes the gyromagnetic factor of the electron to $g_{(1/2)}=2$, but rather the specific form of the primordial quadratic wave equation obeyed by it,  that is amenable to a linearization.
The fact is that space-time symmetries alone  define solely  the kinematic properties of the particles  and  neither fix the values of their interacting constants, nor do they necessarily  prescribe linear Lagrangians. Information on  such properties has to be obtained from additional physical inputs involving  the dynamics. We here provide an example in support of the latter statement. Our case is that the 
spin-$1/2^-$ fermion residing  within the four-vector spinor triad, $\psi_\mu \sim {\Big(}1/2^+$-$1/2^-$-$3/2^-{\Big)}$, whose sectors at the free particle level are interconnected by spin-up and down ladder operators, does not allow for a description within a linear framework at the interacting level. Upon gauging, despite transforming according to the irreducible $(1/2,1)\oplus (1,1/2)$ building block of $\psi_\mu$, and being described by 16 dimensional four-vector spinors, though  of only  four independent components each, its Compton scattering cross sections, both differential and total, result equivalent to those for a spin-$1/2$ particle described by the generalized Feynman--Gell--Mann equation from above (for which we provide an independent algebraic motivation) and with $g_{(1/2^-)}=-2/3$. In effect, the spin-$1/2^-$ particle residing within the four-vector spinor effectively behaves as a true relativistic  ``quadratic'' fermion.
The $g_{(1/2^-)}=-2/3$  value  ensures in addition the desired unitarity in the ultraviolet. In contrast, the spin-$1/2^+$  particle, in transforming irreducibly in the
$(1/2,0)\oplus (0,1/2)$ sector of $\psi_\mu$,   is shown to behave as a truly linear Dirac fermion. Within the framework employed, the three spin sectors of $\psi_\mu$ are described on equal footing by representation- and spin specific wave equations and associated Lagrangians which are of second order in the momenta.
\end{abstract}

\pacs{PACS numbers: 11.30.Cp, 03.65.Pm, 13.40.Em}
\keywords{Lorentz algebra; representation space reduction algorithm;  four-vector spinor; gyromagnetic ratio.}
\maketitle

\tableofcontents

 
 
\section{Introduction}\label{sec1} 
 
The gyromagnetic factor, $g_s$, is one of the fundamental constants characterizing elementary particles as it defines their magnetic moments, 
${ \m} _s$, and thereby the  potential energy,
${\mathcal V}^{s}$, of a particle of spin-$s$
within an external magnetic field, ${\mathbf H}$, given  in the non-relativistic case by,
\begin{equation}
{\mathcal V}^s=-{\mathbf{ \mu}_s}\cdot {\mathbf H},
\quad { \m}_s= g_s \mu_B{\mathbf S},\quad \mu_B=\frac{e\hbar }{2mc}.
\end{equation}
Here, ${\mathbf S}$ stands for the particle's spin,  $m$ is its mass,  $e$ denotes  the electric charge,
$\mu_B$ is the associated elementary magneton, and $g_s$ is the gyromagnetic factor. From now onward the physical quantities will be given in units of $\hbar=1$, $c=1$.
Especially for the electron, this factor takes the value of $g_{(1/2^+)}=2$. This particular $g_s$  value 
is closely related to the linearity of the wave equation describing the electron's  propagation \cite{Greiner}.
{}For a relativistic electron described by means of the Dirac equation coupled to the electromagnetic field,
\begin{equation}
\left(i D^\mu\gamma_\mu -m\right)\psi^D(x)=0,\quad  D^\mu=\partial ^\mu +i{e}A^\mu,\quad \left[D^\mu,D^\nu\right]=ieF_{\mu\nu},
\label{GaugedDirac} 
\end{equation}
where $D^\mu$ is the covariant derivative,  $A^\mu$ is the electromagnetic gauge field, $F_{\mu\nu}$ is the electromagnetic field strength tensor,
while  $\psi^D(x)$ stands for the gauged  Dirac spinor, 
the $g_{(1/2^+)}=2$ value finds a natural explanation upon squaring (\ref{GaugedDirac}). In so doing, one finds
the Klein-Gordon equation coupled to the electromagnetic field in the following particular way,
\begin{eqnarray}
\left( -i\gamma^\mu D_\mu -m\right)\left(i\gamma^\nu D_\nu -m \right)\psi^D(x)=-\left[ \left( \partial_\mu +i{e}A_\mu\right)^2 +2\left(\frac{e}{4}\right)\sigma^{\mu\nu}F_{\mu\nu}+ m^2\right]\psi^D(x)&=&0,
\label{KG_gs2}\\
2\left(\frac{e}{4}\right)\sigma^{\mu\nu}F_{\mu\nu}=-2 \left(\begin{array}{cc}  
 e\frac{\sigma}{2}\cdot \left( {\mathbf H}+i{\mathbf E}\right)&0\\
0& e\frac{\sigma}{2}\cdot\left({\mathbf H}-i{\mathbf E}\right)
\end{array}
\right),  
 \label{tensor_int}
\end{eqnarray} 
implying $g_{(1/2^+)}=2$.
The latter equation shows that the  interacting  Klein-Gordon equation 
can be bi-linearized into the gauged Dirac equation and its conjugate,  only if it has the particular form given in (\ref{KG_gs2}), i.e. exclusively  for
$g_{(1/2^+)}=2$.  The equation (\ref{KG_gs2}) is known under the name of the Feynman-Gell-Mann equation \cite{FGM}.
In a similar way, the linearization of the electromagnetically coupled Schr\"odinger equation  to the Pauli equation leads to  that very same 
$g_{(1/2^+)}$ value \cite{Greiner}.

We here draw attention to the fact that Eq.~(\ref{KG_gs2}) represents a special case of the more general equation,
\begin{eqnarray}
\left[ \left(\partial_\mu +i{e}A_\mu\right)^2 + {g_s}\left(\frac{e}{4}\right)\sigma^{\mu\nu}F_{\mu\nu}+ m^2\right]\psi(x)&=&0,\label{KG_gs22ref}
\end{eqnarray} 
where  $g_s$ can be  any arbitrary real constant, and if $g_s\not=2$, then $\psi (x)\not=\psi^D(x)$. Such an equation, to be referred to as ``generalized Feynman--Gell---Mann equation''  has been obtained in \cite{DelgadoAcosta:2010nx}
along the line of the technique of the Poincar\'e covariant projector method developed in \cite{Napsuciale:2006wr}
from considering the eigenvalue problem of the squared Pauli-Lubanski vector operator, $\o^2(p)$, in the
fundamental representation, $(1/2,0)\oplus (0,1/2)$ of the Lorentz algebra $sl(2,C)\sim so(1,3)$. The $\o_\alpha(p)$ vector expresses as,
\begin{equation}
\o_\alpha(p)=\frac{1}{2}\epsilon_{\alpha\beta \rho\eta}M^{\beta \rho}p^\eta,
\end{equation}
where $M^{\beta \rho}$ stand for the $so(1,3)$ generators in ($1/2,0)\oplus (0,1/2)$, given by,
\begin{equation}
M^{ij}=\frac{i}{4}\left[ \gamma^i,\gamma^j\right]=\frac{1}{2}\sigma^{ij},
\quad M^{0i}=\frac{i}{4}\left[ \gamma^0,\gamma^i\right]=\frac{1}{2}\sigma^{0i}, \quad i,j=1,2,3,
\end{equation}
where $\gamma^\mu$ are the Dirac matrices.
The squared Pauli-Lubanski vector is then calculated as,
\begin{equation}
\o^2(p)= -\frac{1}{4}\sigma_{\lambda\mu} \sigma^\lambda{}_\nu p^\mu p^\nu= 
-\frac{3}{4}p^2 +\frac{1}{4}m \gamma_\eta \gamma_\lambda \left[ p^\lambda, 
p^\eta \right]. 
\label{Pl2} 
\end{equation}
Its eigenvalues on, say, the momentum-space degrees of freedom, $\phi_i(p)  \in (1/2,0)\oplus (0,1/2)$,  are
\begin{equation}
\o^2(p)\phi_i(p)=-m^2\frac{1}{2}\left( \frac{1}{2}+1\right)\phi_i(p), \quad i=1,2,3,4.
\end{equation}
The latter equation is equivalently cast into the form of a covariant spin-$1/2$- and mass $m$  projector,
${\mathcal P}^{(m,1/2)}_{\o^2}(p) $ 
(termed to as Poincar\'e projector)  according to,
\begin{eqnarray}
{\mathcal P}^{(m,1/2)}_{\o^2}(p) \phi_i (p) &=&\phi_i (p),\nonumber\\
{\mathcal P}^{(m,1/2)}_{\o^2}(p)&=&\frac{p^2}{m^2}\left( \frac{\o^2(p)}{-\frac{1}{2}\left(\frac{1}{2}+1\right)p^2}\right).
\end{eqnarray}
In contrast to the standard way, in which one diagonalizes the covariant parity projector, ending up with the Dirac equation,
the $\o^2(p)$ eigenstates are of unspecified parity. However,
because $\o^2(p)$ commutes with the parity operator, the quantum number of parity always can be incorporated into the $i$ label of  $\phi_i(p)$ at a later stage. Same is valid for the polarization index, $\l$.

The corresponding differential equation is then obtained under the replacement, $p_\mu\longrightarrow i\partial_\mu$,
amounting to,
\begin{equation}
\left(\partial^2 -g_s\frac{i}{4}\sigma_{\mu\nu}\left[ \partial ^\mu,\partial^\nu\right] +m^2\right)\psi (x)=0.
\label{KGG1}
\end{equation} 
In the latter, the $so(1,3)$ algebraically prescribed anti-symmetric $\left[ \partial^\mu,\partial^\nu \right]$-piece of the wave equation is weighted  by an arbitrary factor, 
$g_s$, with the motivation that it does not contribute to the free kinematic at all and its weight can not  be uniquely fixed.
 The associated Lagrangian then reads,
\begin{equation}
{\mathcal L}^{(1/2)}_{KG}=\left( \partial^\mu {\bar \psi}(x)\right)
\left(g_{\mu\nu} -g_s\frac{i}{2}\sigma_{\mu\nu}\right)\partial^\nu\psi(x) +m^2{\bar \psi(x) }\psi(x).
\label{KGG2}
\end{equation}
\begin{quote}
The equations (\ref{KGG1}) and (\ref{KGG2}) provide the most general framework for relativistic spin-$1/2$ description in so far as they allow for the existence of truly ``quadratic'' fermions characterized by a gyromagnetic factor $g_{(1/2)}\not=2$.
\end{quote}
This consideration shows that, as already argued in \cite{Greiner}, the $g_{(1/2)}=2$ value for the electron is not due to relativity alone but also to
the form of the quadratic equation obeyed by it, which permits a bi-linearization, be the equation relativistic or not.
The fact that relativity alone does not fix the values of the gyromagnetic ratio has also been noticed in \cite{Napsuciale:2006wr}, where the gyromagnetic factor of 
spin-$3/2^-$ within the four-vector--spinor has been fixed to $g_{(3/2^-)}=2$ from the requirement on causal propagation within an electromagnetic environment, and for the case of a Lagrangian of second order in the momenta (see Appendix II for a more detailed coverage of this issue).

\noindent
The goal of the present study is to provide a consistent and gauge invariant description of the spin degrees of freedom in the
four-vector spinor space, to show that the spin-$1/2^-$ companion
to spin-$3/2^-$ within the ${\Big(}1/2^+$--$1/2^-$--$3/2^-{\Big)}$ triad behaves in Compton scattering as a 
``quadratic'' spin-$1/2$ fermion, and  to explore consequences.

\noindent
The article is structured as follows. In the next section we present first the conventional  complete set of the degrees of freedom (d.o.f.) spanning the four-vector-spinor space and the associated auxiliary conditions through which they are unambiguously identified.
We conclude on the impossibility of a linear description of the spin-$1/2^-$ residing there, and furthermore  notice that both the
spin-$1/2^+$ and spin-$1/2^-$  sectors are reducible with respect to Lorentz transformations. 
Through the text, we denote the reducibility, or the  irreducibility with respect to Lorentz transformations by ``$so(1,3)$ reducibility/irreducibility''.
Motivated by the interconnection of the $\psi_\mu$ degrees of freedom  by spin-up and down ladder operators, we search  
for a possibility to describe all the spin degrees of freedom  on equal footing,  and test  a particular second order formalism,
highlighted in Section 3.  Namely, we present there the recently developed general reduction algorithm  of Lorentz algebra representations \cite{AGK2015} based on the one 
of the momentum independent Casimir invariants of the Lorentz algebra and apply it to the four-vector spinor with the task to split  
the irreducible Dirac sector, $(1/2,0)\oplus (0,1/2)$, from the genuine irreducible Rarita-Schwinger sector, $(1/2,1)\oplus (1,1/2)$. 
Also there,  we perform the separation between the non-interacting d.o.f.'s  of  the spin-$1/2^-$ and $3/2^-$ residents of $(1/2,1)\oplus (1,1/2)$, employing covariant spin-projectors based upon the squared Pauli-Lubanski vector operator, a Casimir invariant of the
Poincar\'e algebra, and present the free particle  wave equations and Lagrangians obtained in this way, which are all representation-- and spin specific, and  of second order in the momenta.
The wave equations  for the spin-$1/2$ four-vector--spinors can be reduced from 16 to 4 degrees of freedom through  contractions either by $p^\alpha$, or $\gamma^\alpha$. The equations obtained in this way  are especially simple, as they appear to be of the type given in (\ref{KGG1}) above.
In Section 4 we couple the  aforementioned \underline{contracted} wave equations to the electromagnetic field
and show that they take each the form of the generalized Feynman--Gell-Mann equation (\ref{KG_gs22ref}).
Section 5 is devoted to the gauged equations and associated Lagrangians in the \underline{full 16 dimensional space}, from which we read off  
the electromagnetic current densities. Furthermore, we  calculate in same space the magnetic dipole moments for spin-$1/2^+$ in $(1/2,0)\oplus (0,1/2)$, 
and spin-$1/2^-$ in $(1/2,1)\oplus (1,1/2)$, and find them fixed to  
$g_{(1/2^+)}=2$, and $g_{(1/2^-)}=-2/3$, respectively. These are the values which also show up in the contracted equations. 
With the aim to check whether the contracted and full-space gauged equations are equivalent, 
 we calculate in section 6 the differential and total  Compton scattering cross sections off spin-$1/2^-$ first with the vertexes from the gauged Lagrangians in the full space. Next we  compare them  with the results alternatively  obtained  with the vertexes following from  the contracted equation, i.e the one that describes this sector in terms of the correct number of four gauged spinorial degrees of freedom, and is shaped after
(\ref{KG_gs22ref}) with $g_{(1/2^-)}=-2/3$.
We encounter an exact coincidence, meaning that the description  of  Compton scattering off this target provided by the
 contracted gauged spin-$1/2^-$ wave equation is indeed  
equivalent to the description provided by  the equation in the full space of the gauged four-vector--spinor degrees of freedom.
This observation allows us first to conclude that the spin-$1/2^-$ resident in $\psi_\mu$ behaves effectively  as a truly ``quadratic'' fermion, and then that
the  suggested  covariant separation prescription of the spin-$1/2^-$ and spin-$3/2^-$ degrees of 
freedom in the irreducible $(1/2,1)\oplus (1,1/2)$ building block of the four-vector--spinor remains valid upon gauging. 
In contrast, the spin-$1/2^+$ is shown to be a truly linear Dirac fermion.  
The paper closes with brief conclusions.  It has one appendix containing the spin-up and down ladder operators within the four-vector spinor space, and a second one
in which we  briefly review for the sake of the self sufficiency of the presentation the here omitted spin-$3/2^-$ degrees of freedom, 
previously extensively studied elsewhere.

\section{The conventional degrees of freedom spanning the four-vector spinor space and their identification 
through auxiliary conditions}
The four-vector spinor, $\psi_\mu$,  is the direct product of the four vector, $ (1/2,1/2)$ and the Dirac spinor,
$(1/2,0)\oplus (0,1/2)$. This representation space  of the Lorentz algebra, $so(1,3)$, is reducible with respect to Lorentz transformations  as, 

\begin{equation}
\psi_\mu\sim \left(\frac{1}{2},\frac{1}{2}\right)\otimes \left[ \left(\frac{1}{2},0\right)\oplus \left(0,\frac{1}{2}\right)\right]\longrightarrow
\left[\left(\frac{1}{2},1\right)\oplus \left(1,\frac{1}{2}\right)\right] 
\oplus \left[ \left( \frac{1}{2},0\right)\oplus 
\left(0,\frac{1}{2}\right)\right].
\label{RS1}
\end{equation}
Accordingly, the number of the degrees of freedom is sixteen, out of which  twelve belong to  $(1/2,1)\oplus (1,1/2)$,
while the remaining four are contained  in $(1/2,0)\oplus (0,1/2)$. If the Dirac particle is considered to be of positive parity, then
the spin and parity content of the first irreducible representation space is $3/2^-$, and $1/2^-$.
Usually, one constructs the degrees of freedom in momentum space, an exercise frequently  performed in the literature, among others in 
\cite{PLB2002}{}, \cite{Napsuciale:2006wr}{}.
 The $(1/2,1/2)$ representation is spanned by one time-like scalar, 
$[\eta_0({\mathbf p},0)]^\a$, 
and three space-like (spin-$1^-$) degrees of freedom,   $[\h({\mathbf p},\ell)]^\a$  (with $\ell =-1,0,1$). 
 The spin-$1/2^+$ four-vector spinors emerging from the coupling of the scalar in 
$(1/2,1/2)$ to the Dirac spinor 
will be termed to as {\underline{\bf S}calar-\underline{\bf S}pinors 
(${\mathbf S}{\mathbf S}$), and denoted by 
$\left[{\mathcal U}^{{\mathbf S}{\mathbf S}}_\pm({\mathbf p},1/2,\l)\right]^\a$, 
finding 
\aeq 
 \left[{\mathcal U}^{{\mathbf S}{\mathbf S}}_{\pm}({\mathbf p},\frac{1}{2}, \l)\right]^\a=
[\eta_0({\mathbf p},0)]^\a 
{u}_{\pm}({\mathbf 
p},\l)=\frac{p^\a}{m} u_{\pm}({\mathbf p},\l).\label{ssdef} 
\ceq 
Here, $[\eta_0({\mathbf p},0)]^\a=p^\a/m$ with $p^2=m^2$, is the only spin-$0^+$ 
vector in $(1/2,1/2)$, where the parity operator is given by the metric tensor $g_{\mu\nu}$, while 
$u_{\pm}({\mathbf p},\l)$ are the usual  Dirac particle-antiparticle spinors in 
$(1/2,0)\oplus(0,1/2)$, 
of positive ($+$) and negative ($-$) parities, respectively, 
(i.e., the $u$-- and $v$-spinors in the terminology of \cite{Bjorken}), and whose 
polarizations are $\l=1/2,-1/2$. 
The spin-$1/2^-$ vector-spinors emerge through the coupling of the spin-$1^-$ four-vectors 
$[\h({\mathbf p},\ell)]^\a$ 
 \cite{Napsuciale:2006wr}, to the Dirac spinor 
and will be termed to  as \underline{\bf V}ector-\underline{\bf S}pinors 
(${\mathbf V}{\mathbf S}$). They are commonly constructed in the standard way by employing  the 
ordinary angular momentum coupling scheme and in terms of appropriate Clebsch- 
Gordan coefficients,  
\begin{subequations}\label{vsdefcg} 
\aec 
~\left[{\mathcal U}^{{\mathbf V}{\mathbf S}}_{\mp}\left({\mathbf p},\frac{1}
{2}, +\frac{1}{2} \right)\right]^\a&=& 
-\sqrt{\frac{1}{3}}[\h({\mathbf p},0)]^{\a}u_{\pm}\left({\mathbf p},+\frac{1}
{2} \right) 
+\sqrt{\frac{2}{3}}[\h({\mathbf p},+1)]^\a u_\pm\left({\mathbf p},-\frac{1}
{2}\right)\label{p12vs},\\ 
~\left[{\mathcal U}^{{\mathbf V}{\mathbf S}}_{\mp}\left({\mathbf p},\frac{1}
{2}, 
-\frac{1}{2}\right)\right]^\a&=& 
\sqrt{\frac{1}{3}}[\h({\mathbf p},0)]^{\a}u_{\pm}\left({\mathbf p},-\frac{1}
{2}\right)-\sqrt{\frac{2}{3}}[\h({\mathbf p},-1)]^\a 
u_\pm\left({\mathbf p},+\frac{1}{2}\right).\label{m12vs} 
\cec 
\end{subequations} 
Notice that if  $[\eta_0({\mathbf p},0)]^\a$ is chosen to be  of positive parity, as done here, the 
parity of $[\h({\mathbf p},\ell)]^\a$ is negative. In consequence, 
the positive- (negative-) parity scalar-spinors are made up of positive- 
(negative-) parity Dirac spinors, while 
the positive- (negative-) parity vector-spinors are made up of negative- 
(positive-) parity Dirac spinors. 
{}Finally,  the spin-$3/2^-$ degrees of 
freedom are given by the following widely spread Clebsch-Gordan combinations, which we 
here list solely for the sake of self-sufficiency of the presentation: 
\begin{eqnarray} 
\left[{\mathcal U}_\mp \left(\mathbf{p},\frac{3}{2},+\frac{3}
{2}\right)\right]^\a &=&\[\h(\mathbf{p},+1)\]^\a 
u_\pm\left(\mathbf{p},+\frac{1}{2}\right),\label{FirstRS}\\ 
\left[{\mathcal U}_\mp\left(\mathbf{p},\frac{3}{2},+\frac{1}
{2}\right)\right]^\a 
&=&\frac{1}{\sqrt{3}}\[\h(\mathbf{p},+1)\]^\a u_\pm\left(\mathbf{p},-\frac{1}
{2}\right) 
+\sqrt{\frac{2}{3}}\[ \h(\mathbf{p},0)\]^\a   
u_{\pm }\left(\mathbf{p},+\frac{1} 
{2}\right),\label{CG1}\\ 
\left[{\mathcal U}_\mp\left(\mathbf{p},\frac{3}{2},-\frac{1}
{2}\right)\right]^\a &=& 
\frac{1}{\sqrt{3}}\[ \h(\mathbf{p},-1)\]^\a u_\pm\left(\mathbf{p},+\frac{1}
{2}\right) 
+\sqrt{\frac{2}{3}}\[ \h(\mathbf{p},0)\]^\a u_{\pm} \left(\mathbf{p},-\frac{1} 
{2}\right),\\ 
\[{\mathcal U}_\mp\(\mathbf{p},\frac{3}{2},-\frac{3}{2}\)\]^\a &=& 
\left[\h(\mathbf{p},-1)\right]^\a u_\pm\left(\mathbf{p},-\frac{1}{2}\right). 
\label{ClebshGord} 
\end{eqnarray} 
They have been extensively studied in the literature beginning with \cite{RS},  
have been highlighted in several textbooks \cite{Weinberg:1995mt}, \cite{Lurie},
and will be as a rule left aside from the main body of the manuscript, though we comment on them in the Apppendix B.
In effect, one first finds the four independent scalar-spinors, 
$\left[{\mathcal U}_\pm^{{\mathbf S}{\mathbf S}} ({\mathbf 
p},1/2, \l)\right]^\alpha$, in (\ref{ssdef}),   and then as 
another set, the four independent 
vector-spinors, $\left[{\mathcal U}^{{\mathbf V}{\mathbf S}}_\pm 
\left({\mathbf p},1/2,\l\right)\right]^\alpha$, 
in (\ref{p12vs})-(\ref{m12vs}) summing up 
to eight (including anti-particles) spin-$1/2^\pm $ states, as it should be. 
Together with  the eight 
independent spin-$3/2^-$ states (as a rule not to be considered here further) the 
total of sixteen independent degrees of freedom in the four-vector-spinor is 
recovered.  
We like to emphasize that the degrees of freedom are interconnected through spin-up and down ladder operators, as summarized in the Appendix I,
a circumstance that strongly suggests  to consider all of them on equal footing.
In the next section we will show that the degrees of freedom  given  in the equations (\ref{ssdef}), (\ref{p12vs}), (\ref{m12vs}), 
behave reducible under Lorentz transformations.

\noindent
The form of the spin-$1/2^-$ vector-spinors can be significantly simplified by means of the Pauli-Lubanski vectors,
${\mathcal W}^\mu(p)$, in the four-vector spinor, on the one side, and  the Pauli-Lubanski vector, $\o^\mu(p)$, in the Dirac-spinor, on the other side.
The general expression for a Pauli-Lubanski vector is given by \cite{Wyborne},
\begin{equation} 
\left[ {\mathcal W}_{\lambda }(p)\right]_{AB}=\frac{1}{2}\epsilon_{\lambda 
\rho\sigma\mu}\left[ M^{\rho\sigma} \right]_{AB}p^\mu, 
\label{gen_PL} 
\end{equation} 
where $M^{\rho\sigma}$ are the generators of the Lorentz algebra in the 
representation space of interest, 
while $A$, and $B$ are the sets of indexes that completely characterize its dimensionality. 
Specifically for the four-vector spinor the generators,  $[M^{\mathbf V}_{\m\n}]_{\a\b}$ and $[M^{\mathbf S}_{\m\n}]$,
 within  the respective Four-\underline {\bf V}ector--,  and the Dirac-\underline{\bf S}pinor building blocks read, 
\aec 
~[M^{\mathbf V}_{\m\n}]_{\a\b}&=&i(g_{\a\m}g_{\b\n}-
g_{\a\n}g_{\b\m}),\label{gensv}\\ 
\left[M^{\mathbf S}_{\m\n}\right]_{ab}&=&\frac{1}{2}\left[\s_{\m\n}\right]_{ab}=\frac{i}{4}
[\g_\m,\g_{\n}]_{ab}\label{genss}, \\
~[M_{\m\n}]_{(\a a),(\b b)}&=&[M^{{\mathbf V}}_{\m\n}] _{\a\b}\delta_{ab}+g_{\a\b} \left[M^{{\mathbf 
S}}_{\m\n}\right]_{ab},
\label{gensrs}
\cec 
with $\g_\m$ being the standard  Dirac  matrices. Notice that the operators in 
\eqref{gensrs} are $(16\times 16)$ matrices, as should be  the generators in 
$\psi_\mu$, and consequently carry  next to the Lorentz indexes, also Dirac spinor 
indexes, $(a,b..)$,  here denoted by small Latin letters. 
With that, ${\mathcal W}_\mu(p)$ expresses 
as the direct sum of the Pauli-Lubanski vectors, $W_\mu(p)$,  and $\o_\mu(p)$, in 
the 
respective  $(1/2,1/2)$- and 
Dirac-building blocks according to \cite{Napsuciale:2006wr}, 
\begin{eqnarray} 
\left[{\mathcal W}_\mu (p)\right]_{(\alpha a)(\beta b)} &=& 
\left[ \o_\mu(p)\right]_{ab} g_{\a_\beta} + 
\left[W_\mu(p)\right]_{\alpha_\beta}\delta_{ab},\nonumber\\ 
\left[\o_\mu (p)\right]_{ab}=-\frac{1}{2}\left[ \gamma_5(p_\mu -\gamma_\mu  p\!\!/)\right]_{ab}, &\quad& 
\left[ W_\mu(p)\right]_{\alpha}{}^{\beta}=i\epsilon_{\mu\alpha} 
{}^\beta{}_{\sigma}p^\sigma. 
\label{PaLu_VS} 
\end{eqnarray} 
In reference to  \eqref{gen_PL}, the label $A$ in the case under consideration  consists of a Lorentz index, $\a$, and a 
spinor 
index, $a$, i.e. $A$ presents itself as a double (four-vector)--(Dirac-spinor) index, 
$A\simeq (\a a)$.
{}Finally, the squared Pauli-Lubanski vector in the four-vector-- spinor is calculated as,
\aec 
~[{\mathcal W}^2 (p)]_{\a\b}&=&[{\mathcal W}^\sigma  (p)]_{\a}{}^\g [{\mathcal 
W}_\sigma  (p)]_{\g\b} 
\label{w2rs}\\ 
&=&\frac{1}{4}\e^{\s}{}_{\l\t\m}\e_{\s\h\z\n}[M^{\l\t}]_{\a} 
{}^\d 
[M^{\h\z}]_{\d\b}p^\m p^\n.\label{tw2rs}
 \cec 
where the Dirac spinor indexes have been suppressed for the sake of simplifying the notations.
The squared Pauli-Lubanski vector operator can be diagonalized within the four-vector spinor space as,
\begin{equation}
\left[{\mathcal W}^2(p)\right]^\a\,_\b\psi^\b=-p^2s_i(s_i+1)\psi^\a, \quad i=1,2,\quad s_1=\frac{1}{2},\quad  s_2=\frac{3}{2}.
\label{W2eigenvalues}
\end{equation}
After some algebra, the following subtle  
relationship between the squared Pauli-Lubanski vector ${\mathcal W}^2(p)$ in Eqs.~(\ref{w2rs})-(\ref{tw2rs}), 
on the one side, and the Pauli-Lubanski vector $\omega_\alpha(p) $ in the Dirac- 
representation  (\ref{PaLu_VS}),
\aeq 
~[{\mathcal W}^2(p)]_{\a}{}^\b \omega_\b  (p) =-p^2\frac{1}{2}\(\frac{1}
{2}+1\)\omega_\a (p), 
\label{w2wsrel} 
\ceq 
can be verified. The latter equation indicates that the spin-$1/2^-$ 
degrees of freedom in $\psi_\mu$  can be equivalently re-expressed  by 
the aid of the Pauli-Lubanski vector, $\omega_\a (p)$, from the Dirac building block. 
Indeed, one can easily verify component by component that the spin-$1/2^-$ 
states  in the equations 
\eqref{p12vs} and \eqref{m12vs}   are equivalently cast as,
\aec 
~\left[{\mathcal U}^{{\mathbf V}{\mathbf S}}_{\pm}\left({\mathbf p},\frac{1}
{2}, \l\right)\right]^\a 
&=&\frac{2} {\sqrt{3}\,m}\omega^\a  (p) \g^5 u_{\pm}({\mathbf p},\l)\nonumber\\ 
&=&\frac{2}{\sqrt{3}\,m} \(\frac{1}{2}\g^5(-p^\a+\g^\a\slashed{p})\)\g^5 
u_{\pm}({\mathbf p},\l)\nonumber\\ 
&=&\frac{1}{\sqrt{3}\,m} \(-p^\a+\g^\a\slashed{p}\)u_{\pm}({\mathbf 
p},\l)\label{vsdef}. 
\cec 
The expressions in \eqref{ssdef} and \eqref{vsdef} will proof very useful in 
the 
following. Specifically, the 
demonstration of the orthogonality between scalar- and vector-spinors will 
benefit from the 
well known property of the Pauli-Lubanski vectors of being divergence-less, 
\aeq 
p^\a \omega_\a  (p) =0, 
\ceq 
which  implies  that  the vector- 
spinors satisfy the following auxiliary conditions: 
\begin{eqnarray}
p^\a \left[{\mathcal U}^{{\mathbf V}{\mathbf S}}_{\pm}
\left({\mathbf p},\frac{1}{2}, \l\right)\right]_\a&=&0,\label{aux1}\\
(p\!\!\!/\pm m) \o ^\a (p)
\left[{\mathcal U}^{{\mathbf V}{\mathbf S}}_{\pm}\left({\mathbf p},\frac{1}{2}, \l\right)\right]_\a &=&0.
\label{aux2}
\end{eqnarray}
However, the first condition is not exclusive to the spin-$1/2^-$ particle but is also shared by the spin-$3/2^-$ degrees of freedom, characterized by
the well known set of auxiliary conditions \cite{RS},\cite{Lurie},
\begin{eqnarray}
p^\a \left[{\mathcal U}_\pm \left(\mathbf{p},\frac{3}{2},\l \right)\right]_\a&=&0,
\label{32aux1}\\ 
\g^\a \left[{\mathcal U}_\pm \left(\mathbf{p},\frac{3}{2},\l \right)\right]_\a &=&0.
\label{32aux2}
\end{eqnarray}
Instead, $\left[{\mathcal U}^{{\mathbf S}{\mathbf S}}_{\pm}({\mathbf p},1/2,\l)\right]^\a$ is uniquely identified by, 
\begin{equation}
{\mathcal W}^\a (p) \left[{\mathcal U}^{{\mathbf S}{\mathbf S}}_{\pm}({\mathbf p},\l)\right]_\a=0, 
\label{auxcnd}
\end{equation}
an observation due to \cite{PLB2002}.
All three different spin-parity sectors obey the Dirac equation:
\begin{equation}
(p\!\!\!/\pm m)\left[{\mathcal U}_\pm \left(\mathbf{p},\frac{3}{2},\l \right)\right]_\a=
(p\!\!\!/\pm m) \left[{\mathcal U}^{{{\mathbf V}{\mathbf S}}}_\pm \left(\mathbf{p},\frac{1}{2},\l \right)\right]_\a=
(p\!\!\!/\mp m) \left[{\mathcal U}_\pm^{{\mathbf S}{\mathbf S}} \left(\mathbf{p},\frac{1}{2},\l \right)\right]_\a=0.
\label{Dirceq}
\end{equation}

In consequence, anyone of the reducible degrees of freedom of the four-vector--spinor satisfies the Dirac equation and 
can be uniquely characterized by its own set of auxiliary conditions \cite{PLB2002}. 
As already announced  in the previous section, from now on we focus 
on the two spin-$1/2^\pm $ sectors in $\psi_\mu$,  summarized in Table 1.
\begin{table} 
\begin{tabular}{||c||c|} \hline 
{$so(1,3)$ reducible free spin-$1/2$ degrees of freedom}  & {Wave equation and auxiliary condition(s)}   \\ 

\hline 
~&~\\
~& 1.\,\, $\left( \slashed{p}\mp m\right) \[{\mathcal U}^{{\mathbf S}{\mathbf S}}_{\pm}\left({\mathbf p}, \frac{1}
{2},\lambda  \right)\]^\a =0$\quad eq.~(\ref{Dirceq})\\
~&~\\ 
spin-1/2$^+$:\,\, $\[{\mathcal U}^{{\mathbf S}{\mathbf S}}_{\pm}\left({\mathbf p}, \frac{1}
{2},\lambda  \right)\]^\a=\frac{p^{\a}}{m}u_{\pm}\left({\mathbf 
p},\lambda  \right)$, \quad eq.~\eqref{ssdef} & \, 2.\,\,  ${\mathcal W}^\a(p) \[{\mathcal U}^{{\mathbf S}{\mathbf S}}_{\pm}\left({\mathbf p}, \frac{1}
{2},\lambda  \right)\]_\a$=0 \quad eq.~\eqref{auxcnd} \\ 
~&~\\ 
\hline 
~&~\\
~&1.\,\, $\left( \slashed{p}\pm m\right)\[{\mathcal U}^{{\mathbf V}{\mathbf S}}_\pm\left({\mathbf p}, \frac{1}
{2},\lambda  \right)\]^\a =0$\quad\,\, eq.~(\ref{Dirceq})\\
~&~\\
spin-$1/2^-$:\,\,$\[{\mathcal U}^{{\mathbf V}{\mathbf S}}_\pm\left({\mathbf p}, \frac{1}
{2},\lambda  \right)\]^\a=\frac{2}{\sqrt{3}m}\o^\a (p)  \gamma^5u_\mp ({\mathbf p},\l )$&
2.\,\, $(p\!\!\!/\pm m)\o ^\a (p) \[{\mathcal U}^{{\mathbf V}{\mathbf S}}_\pm\left({\mathbf p}, \frac{1}
{2},\lambda  \right)\]_\a=0$\\
~&~\\
eqs.~\eqref{vsdefcg}-\eqref{vsdef}&  eq.~\eqref{aux2}\\
~&~\\
~ & 3.\,\, $p^\a \[{\mathcal U}^{{\mathbf V}{\mathbf S}}_\pm\left({\mathbf p}, \frac{1}
{2},\lambda  \right)\]_\a=0$ \quad eq.~(\ref{aux1})\\
~&~\\
\hline
\end{tabular} 
\caption{ \label{table1} Identification of the $so(1,3)$  reducible spin-$1/2^+$ and spin-$1/2^-$  degrees of freedom within the 
four-vector spinor (left column) through the Dirac equation and proper auxiliary conditions (right column).} 
\end{table} 
It is visible from eq.~\eqref{aux2} and the Table 1 that the auxiliary condition identifying the spin-$1/2^-$ sector is of second order in the momenta.
This shows that its degrees of freedom are lacking a description within the linear Rarita-Schwinger  framework.
The linear framework exploited in the description of the highest spin-$3/2^ -$ is anyway widely known to be  plagued by several inconsistency 
problems \cite{Velo:1970ur}. In one of the possibilities such problems have been circumvented within the second order theory of the Poincar\'e covariant projectors developed in 
\cite{Napsuciale:2006wr}, a framework which we partly follow.

\section{The $so(1,3)$ irreducible degrees of freedom in the four-vector--spinor space and a unified  second order description free from auxiliary conditions}

The goal of the current section is to design a description of  all the degrees of freedom spanning the four-vector spinor space on equal footing.
This could be of interest in studying in physical processes  possible interferences between the highest  spin-$3/2^-$,  with the lower-spin components 
(see \cite{Sholten} for a related discussion). 
Such  turns out to be realizable  within a second order formalism, and with the special emphasis on the strictly  $so(1,3)$ irreducible d.o.f. 
As we shall see in due places, the  approach will be free from auxiliary conditions.
The section is structured as follows. In the first subsection we highlight a recently developed $so(1,3)$ representation reduction algorithm \cite{AGK2015} 
based on static projectors constructed from one of the Casimir invariants of the Lorentz algebra. In the second subsection we identify the $so(1,3)$  irreducible
$(1/2,0)\oplus (0,1/2)$ sector in the four-vector spinor and split it neatly from $(1/2,1)\oplus (1,1/2)$, which now exclusively hosts the
spin-$1/2^-$ and $3/2^-$. In the third subsection we construct along the line of ref.~\cite{Napsuciale:2006wr} covariant mass-$m$ and spin-$1/2$ projectors 
from the Casimir invariants of the Poincar\'e algebra, i.e. from the squared four-momentum, $P^2$, and the squared Pauli-Lubanski vector operator, 
${\mathcal W}^2(p)$, which are of second order in the momenta. Finally, in the fourth subsection we combine the aforementioned  Lorentz- and Poincar\'e  projectors
and obtain the second order wave equations satisfied by the two spin-$1/2$ degrees of freedom from  the irreducible $so(1,3)$ sectors in the four vector spinor.

\subsection{Recognizing  $so(1,3)$ irreducible representation spaces by static projectors  derived from an
invariant  of the Lorentz algebra}\label{sec2b} 
 To ensure that each one of the two  spin-$1/2^ + $ and spin-$1/2^- $ degrees of freedom  of   $\psi_\mu$ 
transforms 
exclusively according to  one of its two $so(1,3)$ irreducible sectors, 
i.e. to  either  $(1/2,0)\oplus (0,1/2)$, or, $(1/2,1)\oplus (1,1/2)$, and thereby to pay a tribute to Wigner's particle definition \cite{Wigner},
it is necessary to introduce projectors based upon one of the Casimir 
invariants 
of the Lorentz algebra. 
The Lorentz algebra has a Casimir operator, denoted by $F$,  and given by \cite{Wyborne},
\aec 
~[F]_{\a\b}&=&\frac{1}{4}[M^{\m\n}]_\a{}^\g 
[M_{\m\n}]_{\g\b}=\frac{9}{4}g_{\a\b}+\frac{i}{2}\s_{\a\b},
\label{Foprtr}
\cec 
where use has been made of 
\eqref{genss}--\eqref{gensv}, and  \eqref{gensrs} to work out the explicit expression on the rhs in (\ref{Foprtr}). 
Its  eigenvalue problem reads, 
\aec 
F \,|j_1,j_2\rangle &=&\frac{1}{2}(K(K+2)+M^2)|j_1,j_2\rangle
=\frac{1}{2}(j_1(j_1+1)+j_2(j_2+1)) |j_1,j_2\rangle, 
\label{FCasm} 
\cec 
where $|j_1,j_2\rangle $ stand for some generic  states 
transforming irreducibly under $so(1,3)$ as $(j_2,j_1)\oplus(j_1,j_2)$, while
\aeq 
K=j_1+j_2,\qquad M=\vert j_1-j_2 \vert. 
\ceq 
The $K$- and $M$-values fully characterize any $(j_2,j_1)\oplus (j_1,j_2)$ 
representation space, remain same in all inertial frames, and allow, if preferred,  for the following relabeling, 
\begin{equation} 
(j_2,j_1)\oplus (j_1,j_2)\simeq (K,M) . 
\label {KMQM} 
\end{equation}
On the basis of $F$ one can construct projector operators on the irreducible sectors,  $(1/2,j_1)\oplus (j_1,1/2)$, in $\psi_\mu$, with $j_1=0, 1$.
Such projectors will be denoted hereafters by, $\mathcal{P}_F^{(1/2,j_1)}$, and  we find them as,
\begin{eqnarray}
\mathcal{P}^{(1/2,0)}_F&=&\(\frac{F-\l_{1}}{\l_{0}-\l_{1}}\),
\label{FDir}\\
\mathcal{P}^{(1/2,1)}_F&=&\(\frac{F-\l_{0}}{\l_{1}-\l_{0}}\),
\label{FRS}
\end{eqnarray} 
where, $\l_{j_1}$ are the $F$ eigenvalues,
\aeq\label{evF} 
\l_0=\frac{3}{4},\qquad 
\l_1=\frac{11}{4}, 
\ceq 
corresponding to $j_2=1/2$ and $j_1=0,1$, respectively.
Their respective momentum-space eigenstates, denoted by $w^{(1/2, j_1)}({\mathbf 
p},s_i, \lambda)$, 
transform accordingly as  the $(1/2,j_1)\oplus(j_1,1/2)$ sectors of the 
four-vector--spinor  representation space: 
\aec 
j_1=0,\quad \left(\frac{1}{2},0\right)\oplus \left(0,\frac{1}{2}\right):&\quad& 
\mathcal{P}^{(1/2,0)}_Fw^{(1/2,0)}\left({\mathbf p},\frac{1}{2},\l \right) 
=w^{(1/2,0)}\left({\mathbf p},\frac{1}{2},\l \right), 
\label{feq12}\\ 
j_1=1,\quad \left(\frac{1}{2},1\right)\oplus \left(1,\frac{1}{2}\right):&\quad& 
\mathcal{P}^{(1/2,1)}_Fw^{(1/2,1)}\left({\mathbf p},\frac{1}{2},\l \right) 
 =w^{(1/2, 1)}\left({\mathbf p},\frac{1}{2},\l \right). 
\label{feq32} 
\cec 
 The respective Lorentz projectors   are then 
easily calculated as, 
\aec 
~[\mathcal{P}^{(1/2,0)}_F]_{\a\b}&=&\frac{1}{4}\g_\a \g_\b,\label{pf0}\\ 
~[\mathcal{P}^{(1/2, 1)}_F]_{\a\b}&=&g_{\a\b}-\frac{1}{4}\g_\a 
\g_\b\label{pf1}. 
\cec 
The orthogonality and completeness properties of the above set of operators 
are easily verified. The Lorentz projectors  have the 
advantage to be  momentum-independent, 
which allows one any time when they are put at work,  to flawlessly recognize within a $so(1,3)$  reducible representation space
its $so(1,3)$ irreducible sectors for which they have been specifically designed, and to 
remove the rest  without 
increasing the power of the momentum dependence  of the wave equation. 
The next section is devoted to the solution of the above eqs.~\eqref{feq12} 
and \eqref{feq32}. 
 
 

 \subsection{Identifying the  irreducible  degrees of freedom in 
\texorpdfstring{$\psi_\mu$}{} and 
equations of motion\label{sec2c}} 
 

It is straightforward to verify that \underline{none} of the 
$\left[{\mathcal U}^{{\mathbf S}{\mathbf S}}_\pm({\mathbf p},1/2, \l)\right]^\alpha 
$ 
and  $\left[ {\mathcal U}^{{\mathbf V}{\mathbf S}}_\pm\left({\mathbf 
p},1/2,\l\right)\right]^\alpha $ four-vector spinors commonly used  as 
bases within the four-vector spinor space and given 
in the respective Eqs.~(\ref{ssdef}), (\ref{p12vs}), and (\ref{m12vs})  from 
above behaves irreducibly under Lorentz 
transformations, this because none of these states acts as  an eigenstate  to 
the $F$--Casimir invariant of the Lorentz algebra in \eqref{FCasm}. 
Towards our goal of finding the explicit expressions for the $so(1,3)$ irreducible 
spin-$1/2^+$ and $1/2^-$  states, from now onwards  denoted by 
$[w^{(1/2, j_1)}_\pm(\mathbf{p},1/2,\l)]^\a$, with the low case index, 
$\pm$, specifying their respective parities (to become indicative of particle-antiparticle upon quantization) 
we  benefit from our knowledge on $\left[{\mathcal U}_\pm^{{\mathbf S}{\mathbf 
S}}({\mathbf p},1/2, \l)\right]^\alpha$ in 
\eqref{ssdef} and 
$\left[ {\mathcal U}^{{\mathbf V}{\mathbf S}}_\pm({\mathbf p},1/2, 
\l)\right]^\alpha$ in \eqref{vsdef} and find their respective $j_1=0$- 
projections, i.e. the projections on $(1/2,0)\oplus (0,1/2)$, in
employing the operator from (\ref{pf0}) according to, 
\aec 
~[\mathcal{P}^{(1/2, 
0)}_F]^{\a\b}\left[{\mathcal U}_\pm^{{\mathbf S}{\mathbf 
S}}\left(\mathbf{p},\frac{1}{2}, \l\right)\right]_\b 
&=&\frac{1}{4 m}\g^\a \slashed{p} u_{\pm}\left(\mathbf{p},\l\right), 
\label{mark1}\\ 
~[\mathcal{P}^{(1/2, 0)}_F]^{\a\b}\left[ 
{\mathcal U}^{{\mathbf V}{\mathbf S}}_\pm \left(\mathbf{p},\frac{1}{2}, 
\l\right)\right]_\b 
&=&\frac{\sqrt{3}}{4 m}\g^\a \slashed{p}u_{\pm}\left(\mathbf{p},\l\right). 
\label{mark2} 
\cec 
In this manner the $(1/2,0)\oplus (0,1/2)$ components of the spin-$1/2^+$ Clebsch- 
Gordan combinations, 
$\left[{\mathcal U}^{{\mathbf S}{\mathbf S}}_\pm({\mathbf p},1/2, \l)\right]^\a$, 
 and $\left[{\mathcal U}^{{\mathbf V}{\mathbf S}}_\pm ({\mathbf p},1/2,\l)\right]^\alpha $, are unambiguously 
identified modulo a normalization constant.
 In a way similar, their  $(1/2,1)\oplus (1,1/2)$ components are identified by 
employing the operator in (\ref{pf1}) as,
\begin{eqnarray} 
~[\mathcal{P}^{(1/2, 1)}_F]^{\a\b}\left[ 
{\mathcal U}_\pm^{{\mathbf S}{\mathbf S}}\left( 
\mathbf{p},\frac{1}{2}, \l\right) 
\right]_\b 
&=&\frac{1}{m}\(p^\a- \frac{1}
{4}\g^\a\slashed{p}\)u_{\pm}\left(\mathbf{p},\l\right),\\ 
~[\mathcal{P}^{(1/2, 1)}_F]^{\a\b}\left[ 
{\mathcal U}^{{\mathbf V}{\mathbf S}}_\pm\left( 
\mathbf{p},\frac{1}{2},\l\right)\right]_\b 
&=&-\frac{1}{\sqrt{3}m}\(p^\a-\frac{1}{4}\g^\a 
\slashed{p}\)u_{\pm}\left(\mathbf{p},\l\right). 
\end{eqnarray} 
Taking care of the  normalizations, the non-interacting irreducible spin-$1/2^+$ states, $[w_\pm^{(1/2, j_1)}
(\mathbf{p}, 1/2,\l)]^\a$,  
can finally be cast in terms of covariant entities of factorized Dirac-and 
four-vector degrees of freedom, 
the former being given by the ordinary Dirac spinors, and the latter by 
$p^\a$, and/or $\gamma^\a$, 
 \aec 
\mbox{\footnotesize spin}-\frac{1}{2}^+\in \left(\frac{1}{2},0\right)\oplus \left(0,\frac{1}{2}\right):\quad ~\left[w_\pm^{(1/2,0)}\left(\mathbf{p},\frac{1} 
{2},\l\right)\right]^\a&=&\frac{1}{2 m}\g^\a \slashed{p} 
u_{\pm}\left(\mathbf{p},\l\right)\label{defl0s}\nonumber\\
&=&\pm\frac{1}{2}\gamma^\a u_\pm 
({\mathbf p},\l),\\ 
\mbox{\footnotesize spin}-\frac{1}{2}^-\in \left(\frac{1}{2},1\right)\oplus \left(1,\frac{1}{2}\right):\quad ~\left[w_\pm^{(1/2,1)}\left(\mathbf{p},\frac{1} 
{2},\l\right)\right]^\a&=&\frac{2}{\sqrt{3} 
m}(p^\a-\frac{1} 
{4}\g^\a\slashed{p})u_{\pm}\left(\mathbf{p},\l\right),
\label{defl1s} 
\cec 
where use has been made of the Dirac equation. 
Therefore, one finds  two polarizations  available for 
each parity and two possible parities for each $j_1$ value, making a total of 
eight spin-$1/2$ independent states which now reside each in two  
distinct irreducible  Lorentz invariant representation subspaces of the four-vector--spinor. 
The  above eigenstates to the Lorentz projector are also of well defined parities, and are normalized  as, 
\aeq 
~\left[\overline{w}_\pm^{(1/2, j_1)}\left(\mathbf{p},\frac{1}{2}, 
\l\right)\right]^\a\left[w_\pm^{(1/2, 
j_1)}\left(\mathbf{p},\frac{1}{2},\l\right)\right]_\a=\pm 1, 
\label{ppdef} 
\ceq 
 their  conjugates being defined as, 
\aeq 
\overline{w}_\pm^{(1/2, j_1)}\left(\mathbf{p},\frac{1}{2},\l\right)= 
\left[\g^0 {w}_\pm^{(1/2,j_1)}\left(\mathbf{p},\frac{1}{2}, 
\l\right)\right]^\dagger. 
\label{ppd} 
\ceq 
{}Finally, the highest spin-$3/2^-$ degrees of freedom in the Eqs.~
(\ref{CG1})--(\ref{ClebshGord}) are automatically $so(1,3)$ irreducible because they 
transform exclusively according $(1/2,1)\oplus (1,1/2)$  and imply, 
\begin{eqnarray} 
\left[w^{(1/2,1)}_\pm \left({\mathbf p}, \frac{3}{2},\sigma 
\right)\right]^\alpha &=& \[{\mathcal U}_\pm\left({\mathbf p},\frac{3} 
{2},\sigma\right)\]^\a.
\label{32_irrd} 
\end{eqnarray}

\subsection{The covariant mass-\texorpdfstring{$m$}{} and spin-\texorpdfstring{$1/2$}{} projector from the 
invariants of the Poincar\'e algebra }\label{sec2a} 
 

As repeatedly emphasized, the $so(1,3)$ irreducible Rarita-Schwinger sector $(1/2,1)\oplus (1,1/2)$  contains  
two spins, namely $s_1=1/2^-$ and $s_2=3/2^-$, and one can construct 
covariant mass-$m$ and  spin-$s_i$  projectors, $\mathcal{P}^{(m,s_i)}_{{\mathcal W}^2}(p)$,  
in terms of the Casimir invariants of the Poincar\'e algebra, the squared four-momentum, $P^2$,
and the squared Pauli-Lubanski vector,  ${\mathcal W}^2(p)$ whose eigenvalues are given in the equation (\ref{W2eigenvalues}). 
Such can be done following  the 
prescription given in \cite{Napsuciale:2006wr} leading to, 
\aec 
\left[ \mathcal{P}^{(m,1/2)}_{{\mathcal W}^2}(p)\right]^{\alpha\beta} \, \
&=&\frac{P^2}{m^2}\left[ \(\frac{{\mathcal W}^2(p)-\e_{3/2}}{\e_{1/2}- 
\e_{3/2}}\)\right]^{\alpha\beta}\,, \label{w2eq12}\\ 
\left[ \mathcal{P}^{(m,3/2)}_{{\mathcal W}^2}(p)\right]^{\alpha\beta}\, 
 &=&\frac{P^2}{m^2}\left[ \(\frac{{\mathcal W}^2(p)  -\e_{1/2}}{\e_{3/2}- 
\e_{1/2}}\)\right]^{\alpha\beta}.\label{w2eq32} 
\cec 
Here, $\e_{s_i}=-p^2 s_i(s_i+1)$, with  $s_1=1/2$, and $s_2=3/2$, is the ${\mathcal W}^2(p)$ eigenvalue 
corresponding to the mass-$m$ and spin-$s_i$ eigenstates of the operators 
$P^2$ and ${\mathcal W}^2(p)$. 
Defining now  the tensor $\Gamma_{\m\n}^{(m,1/2)}$ according to,
\aec 
\left[\mathcal{P}^{(m,1/2)}_{{\mathcal W}^2}(p)\right]_{\a\b} 
&=&\frac{1}{m^2} 
\left[\Gamma^{(m,1/2)}_{\m\n}\right]_{\a\b}p^\m p^\n =\frac{p^2}{m^2} 
\left[ 
\mathbb{P}^{(1/2)}(p) 
\right]_{\a\b}, 
\label{pro_12} 
\cec 
where $\left[\mathbb{P}^{(1/2)}(p)\right]_{\a\b}$ is the covariant spin-$1/2$ projector \cite{VanNieuwenhuizen:1981ae},
\aec 
\left[\mathbb{P}^{(1/2)}(p) \right]_{\a\b} 
&=&\frac{1}{3}\g_\a \g_\b+\frac{1}{3p^2}(\slashed{p}\g_\a p_\b+p_\a \g_\b 
\slashed{p})\label{p12}, 
\cec 
one arrives at,

\begin{eqnarray}
\left[\Gamma^{(m,1/2)}_{\mu\nu}\right]_{\alpha\beta}&=&\frac{1}{6 } 
\(\s_{\a\m}\s_{\b\n}+\s_{\a\n}\s_{\b\m}+ 
i\s_{\m\n}g_{\a\b}+5g_{\a\n}g_{\b\m}+g_ {\a\m}g_{\b\n}\). 
\label{Gammam12}
\end{eqnarray} 
The spin $3/2^-$ projector, $\left[\mathcal{P}^{(m,3/2)}_{{\mathcal W}^2}(p)\right]_{\a\b}$ is then obtained as the difference between the unit operator
in $\psi_\mu$  and $\left[\mathcal{P}^{(m,1/2)}_{{\mathcal W}^2}(p)\right]_{\a\b}$. It has been elaborated in detail in 
\cite{Napsuciale:2006wr} and we limit ourselves to briefly present it in the Appendix II.  
 
\subsection{Full-space, and contracted wave equations for the $so(1,3)$ irreducible spin-\texorpdfstring{$1/2^+$}{} and spin-\texorpdfstring{$1/2^-$}{} states in \texorpdfstring{$\psi_\mu$}{}}

So far we have been constructing the explicit degrees of freedom and have read off from their forms the wave equations and the 
conditions they obey, given in the Table 1. In the current subsection we focus on the inverse problem, namely, we shall be seeking for a 
general method of identifying the irreducible degrees of freedom  in the four-vector--spinor and their  wave equations,
from which the explicit forms listed in the previous Section 2 can be obtained as solutions.  
Though we  shall be keeping same notations as above,  no use of the explicit expressions will be made.
In order to find the equations of motion satisfied by the irreducible spin-$1/2^+$ and spin-$1/2^-$ spinors,
the following eigenvalue problem needs to be solved,
\aec 
\Pi _{(1/2,j_1)\oplus(j_1,1/2)}(p)w_\pm^{(1/2,j_1)}\left({\mathbf p},\frac{1} 
{2},\lambda\right)&=&w_\pm^{(1/2,j_1)}\left({\mathbf p},\frac{1} 
{2},\lambda\right), \quad j_1=0,1,
\nonumber\\ 
 \Pi _{(1/2,j_1)\oplus(j_1,1/2)}(p)&=& 
\mathcal{P}^{(1/2, j_1)}_F\,\mathcal{P}^{(m,1/2)}_{{\mathcal W}^2}(p)= 
\mathcal{P}^{(m,1/2)}_{{\mathcal W}^2}(p)\,\mathcal{P}^{(1/2, j_1)}_F, 
\label{comeq} 
\cec 
with $\Pi _{(1/2,j_1)\oplus(j_1,1/2)}(p)$, to be termed to as Lorentz- and Poincar\'e invariant projectors,  being the products 
of the covariant-mass-$m$ and spin-$1/2$ projector from 
Eqs.~(\ref{w2eq12}), on the one side, 
with the respective Lorentz projectors on $(1/2,0)\oplus (0,1/2)$, and 
$(1/2,1)\oplus (1,1/2)$ from Eqs.~(\ref{feq12}), and (\ref{feq32}), on the 
other. Though the projector used, strictly speaking, does not fix the parity of the states, i.e. it does not distinguish between
low case plus and minus indexes, and also does not fix the polarizations, $\lambda$, we here nonetheless shall stick to the full notation, to
be  used through the paper, for the purpose of not unnecessarily changing notations.
This with the justification, that the states of fixed parities and polarizations at any rate solve the equation (\ref{comeq}) 
(with the Lorentz-- and Poincar\'e- projectors from the respective eqs.~(\ref{pf0})-(\ref{pf1})), and (\ref{pro_12})) as,  
 \aeq 
~[\G^{(1/2,j_1)}_{\m\n}]_{\a\b}p^\m p^\n= m^2 [\mathcal{P}^{(1/2, 
j_1)}_F]_{\a}{}^\g [\mathcal{P}^{(m,1/2)}_{{\mathcal W}^2}(p)]_{\g\b} 
=[\mathcal{P}^{(1/2, j_1)}_F]_{\a}{}^\g[\Gamma^{(m,1/2)}_{\m\n}]_{\g\b}p^\m 
p^\n , \quad j_1=1,
\label{Victor}
\ceq 
where $[\Gamma^{(m,1/2)}_{\m\n}]_{\g\b}$ has  been previously defined in 
(\ref{Gammam12}). The relevant kinetic tensor, $\left[\Gamma^{(1/2,j_1)}_{\mu\nu} \right]_{\a\b}p^\mu p^\nu $, defining the wave equations 
for both the spin-$1/2^+$ and $1/2^-$ ,  and given in the latter equation (\ref{Victor}),
contains several anti-symmetric $\left[ \partial_\mu , \partial_\nu \right]$ terms which are identically vanishing at 
the free particle level. In picking up some of those terms and dropping others, several equivalent non-interacting equations can be obtained,
which however will become distinguishable upon gauging, only one of them for each sector, being the physical. 
We will make our choices as explained below and in terms of the following short hands of orthogonal $(4\times 4)$ matrices in 
the Dirac spinor indexes, 
\aec 
~[f^{(1/2, 0)}(p)]^\a&=&\frac{1}{2m}\g^\a \slashed{p},\label{ef0}\\ 
~[f^{(1/2, 1)}(p)]^\a&=&\frac{2}{\sqrt{3}m}(p^\a-\frac{1} 
{4}\g^\a\slashed{p})\label{ef1}, 
\cec 
 orthonormalized on the mass-shell according to, 
\aeq\label{eqforty} 
~[\overline{f}^{(1/2, j)}(p)]^\a[f^{(1/2, j')}(p)]_\a=\d^j_{j'}\frac{p^2} 
{m^2},
\ceq 
with 
$[\overline{f}^{(1/2, j_1)}(p)]^\a=\g^0([f^{(1/2, j_1)}(p)]^\a)^\dagger\g^0$. 
The $[f^{(1/2,j_1)}(p)]^\a$ matrices bring the great advantage to  bi-linearize the kinetic terms of the 
equations of motion (\ref{comeq}) as, 
\aeq\label{fequations} 
\left(m^2[f^{(1/2, j_1)}(p)]_\a [\overline{f}^{(1/2, j_1)}(p)]_\b - 
m^2g_{\a\b}\right)\left[w^{(1/2, j_1)}_\pm \left({\mathbf p},\frac{1} 
{2},\l\right)\right]^\b =0. 
\ceq 
 {}Factorizing the  momenta, the equation (\ref{fequations})  translates into, 
\aeq\label{eomsj1} 
([{\widetilde \G}^{(1/2,j_1)}_{\m\n}]_{\a\b}p^\m p^\n-m^2 g_{\a\b})\left[w^{(1/2, 
j_1)}_\pm\left({\mathbf p},\frac{1}{2},\lambda \right)\right]^\b =0, 
\quad j_1=0,1,
\ceq 
where new tensor $[{\widetilde \G}^{(1/2,j_1)}_{\m\n}]_{\a\b}$ will be the one to be used systematically in the following.\\

\noindent 
\underline{Spin-$\frac{1}{2}^+$:}\\

\noindent
The spin-$1/2^+$ wave equation following from (\ref{ef0}) and (\ref{fequations}) reads, 
\aeq 
\(\frac{1}{4}\left( \g_\a \slashed{p}\slashed{p}\g_\b\right)_{ab} -m^2\delta_{ab} g_{\a\b}\)\left[w^{(1/2, 
0)}_\pm\left({\mathbf p},\frac{1}{2},\lambda\right)\right]^{\b b}=0,
\label{dir_12} 
\ceq
where we as an exception and temporarily brought back the Dirac indexes, $a,b$ for the sake of rising the transparency of the discussion on the solutions to follow.
Notice that this is a genuine quadratic and relativistic fermion equation, not a scalar one. The information on the
dimensionality of the four-vector-spinor space  has fully been encoded by the Lorentz indexes of the gamma matrices and their Dirac indexes.

In order to find out the solutions, we notice that the Dirac index of  $ \left[ w^{(1/2,0)}_\pm\left({\mathbf p}, 1/2, \l\right)\right]_{\a  b}$
has to be carried by a  spinor, $\left[u_\pm ({\mathbf p},\lambda)\right]_b$,
while the Lorentz index can be carried either by
$p_\a\left(\frac{\slashed{p}}{m} \right)^{n_1}$, or by   $\gamma_\a\left( \frac {\slashed{p}}{m}\right)^{n_2}$, for $n_1\geq 0$, and $n_2\geq 0$. 
The first solution is ruled out for not being an ${\mathcal P}_F^{(1/2,0)}$ eigenstate, while  the second satisfies (\ref{dir_12}), and one can chose the lowest 
$n_2=0$ value. In this way, the explicit form previously found in (\ref{defl0s}) is recovered modulo a normalization constant.
In consequence,  the four-vector spinors describing the irreducible Dirac sector are such that their contractions with the
$\gamma$-matrices behave as solutions to the Dirac equation, meaning,
 \begin{eqnarray} 
\gamma^\a _{ab} \left[w^{(1/2,0)}_\pm\left({\mathbf p}, \frac{1}{2}, \l\right)\right]_{\a b} &\sim & 
\left[u_\pm ({\mathbf p},\l)\right]_a,\nonumber\\
(\slashed{p}\mp m)\left[ 
\gamma^\a _{ab} \left[ w^{(1/2,0)}_\pm\left({\mathbf p}, \frac{1}{2}, \l\right)\right]_{\a b}\right]&=&0.
\label{glory} 
\end{eqnarray} 

Finally, contraction of (\ref{dir_12}) by $\g^\a$ from the left yields, 
\aeq 
\( \slashed{p}\slashed{p}\delta_{ab}-m^2\delta_{ab}\)\left[\gamma \cdot w^{(1/2,0)}_\pm \left({\mathbf p}, \frac{1}{2}, 
\l\right)\right]_b = 0.
\label{j0cfact} 
\ceq 
Taking into account that,
\begin{equation}
\slashed{p}\slashed{p} =p^2 -i\frac{\sigma^{\mu\nu}}{2}\left[p_\mu,p_\nu \right],
\label{one}
\end{equation}
one arrives at,
\begin{eqnarray}
\( p^2-i\frac{g_{(1/2^+)}}{4}\sigma^{\mu\nu}\left[p_\mu,p_\nu\right] -m^2\)\left[\gamma \cdot w^{(1/2,0)}_\pm \left({\mathbf p}, \frac{1}{2}, 
\l\right)\right]&=&0,\label{newtree}
 \end{eqnarray}
with $g_{(1/2^+)}=2$, to be independently confirmed below.
The  equation (\ref{glory})  shows that $\left[\gamma\cdot w^{(1/2,0)}_{\pm}({\mathbf p}, 1/2,\l)\right]$ satisfies the Dirac equation and behaves as 
a free spin-$1/2^+$ Dirac particle, while (\ref{j0cfact}) confirms that each spinor component satisfies the free Klein-Gordon equation, as it should be
for a free particle on on mass-shell. \\

\noindent
\underline{Spin-$\frac{1}{2}^-$:}\\

\noindent
The emerging  wave equation relevant for this case  can be cast as, 
\aeq 
\[\frac{4}{3}\(p_\a\delta_{ab}-\frac{1}{4}\left[\g_\a\right]_{ab}\slashed{p}\)\(p_\b\delta_{ab}-\frac{1} 
{4}\slashed{p}\left[\g_\b\right]_{ab}\)-m^2 
\delta_{ab}g_{\a\b}\]\left[ w^{(1/2, 1)}_\mp \left({\mathbf p},\frac{1} 
{2},\l\right)\right]^{\b b}=0.\label{ceq1} 
\ceq 
The latter incorporates the following  property, revealed  through 
its contraction  by $\g^\a$,  
\aeq 
-m^2\left[\g_\b\right]_{ab} \left[w_\pm^{(1/2, 1)}\left({\mathbf p},\frac{1} 
{2},\lambda\right)\right]^{\b b}=0.\label{auxj11} 
\ceq 
It needs to be emphasized that within the second order formalism advocated here, 
the latter equation does not play the role of an auxiliary condition. The principle  equation (\ref{ceq1}) 
is \underline{free from auxiliary conditions}, and
the equation (\ref{auxj11}) reflects only one of its properties.  
The meaning of \eqref{auxj11} is that  $w^{(1/2,1)}_\pm \left({\mathbf 
p},1/2,\lambda\right)$ does not have any 
projection on 
$(1/2,0)\oplus(0,1/2)$, as it should be, and in accord with the 
$so(1,3)$ reduction of 
the Rarita-Schwinger space  established in (\ref{RS1}). 
The Dirac index, $b$,  of the  $\left[ w^{(1/2,1)}_\pm \left({\mathbf p}, 1/2, \l\right)\right]_{\b b}$ spinor has to be carried by a  spinor, while the Lorentz index, $\beta$,
can be carried equally well by the four momentum, $p_\b$, on the one side, and by the Dirac matrices,
$\gamma_\b$, on the other, ending up by  a linear combination of them,
\begin{equation}
\left[ w^{(1/2,1)}_\pm \left({\mathbf p},\frac{1}{2},\l \right)\right]_{\b b} =
c_1 p_\b \delta_{ab}\left[u_\pm ({\mathbf p},\l )\right]_b +c_2\left[\gamma_\b\right]_{ab} \left[u_\pm ({\mathbf p},\l)\right]_{b},
\quad c_2=-\frac{1}{4}c_1.
\label{newtwo}
\end{equation}
The coefficients are determined from the requirement that the states also diagonalize the
Lorentz projector, ${\mathcal P}_F^{(1/2,1)}$, in combination with their normalizations.
In this way, the explicit form previously known by construction  from 
(\ref{defl1s}), can be recovered, yielding,
\begin{eqnarray} 
\left[p\cdot w^{(1/2,1)}_\pm \left({\mathbf p},\frac{1}{2},\l 
\right)\right]_a&\sim &\left[  u_\mp  ({\mathbf p},\l)\right]_a,\nonumber\\
(\slashed{p}\pm m)\left[p\cdot w^{(1/2,1)}_\pm \left({\mathbf p},\frac{1}{2},\l 
\right)\right]_a &=&0. 
\label{dirgdw2} 
\end{eqnarray} 
In this fashion, one  confirms that also $p\cdot w^{(1/2,1)}_\pm ({\mathbf p},1/2,\lambda )$  
is described in terms of an ordinary Dirac equation and therefore behaves as a free spin-$1/2^-$ . \\

{}Finally, contraction of (\ref{ceq1}) by $p^\alpha $ from the left yields,

\begin{eqnarray} 
\left[ \frac{4}{3}\left(p^2-\frac{1}{4}\slashed{p}\slashed{p}\right) -
m^2\right] 
\left[p\cdot w^{(1/2,1)}_\pm \left({\mathbf p},\frac{1}{2},\l \right)\right] 
-\frac{1}{3}\left(p^2-\frac{1}{4}\slashed{p}\slashed{p}\right)\slashed{p} 
\left[\gamma \cdot w^{(1/2,1)}_\pm \left({\mathbf p},\frac{1}{2},\l 
\right)\right]&=&0, 
\label{dirgdw1} 
\end{eqnarray} 
where the last term can be ignored by virtue of the property  
in (\ref{auxj11}).
In effect, the  wave equation relevant for the case under consideration assumes the form,

\begin{equation}
\left( p^2-\frac{1}{4}\slashed{p}\slashed{p} -\frac{3}{4}m^2\right)\, \left[p\cdot w^{(1/2, 1)}_\pm \left({\mathbf p},\frac{1}
{2},\l\right)\right]=0.
\label{eco}
\end{equation}
Taking into account (\ref{one}), 
and factorizing $3/4$, the equation (\ref{eco}) equivalently rewrites to,

\begin{equation}
\left( p^2 -i\frac{g_{(1/2^-)}}{4}\sigma^{\mu\nu}\left[p_\mu,p_\nu \right] -m^2\right)\left[ p\cdot w_\pm ^{(1/2,1)}({\mathbf p},\frac{1}{2},\l)\right],
\label{newfour}
\end{equation} 
with  $g_{(1/2^-)}=-2/3$, to be independently confirmed below.
As a reminder, at the free particle level, the presence of the $[p^\mu, p^\nu]$ term is irrelevant and the equation is a pure scalar one, in which the information of the dimensionality of the $so(1,3)$ representation space is completely lost. However, this  changes upon gauging, when
the term becomes viable by virtue of 
\aeq 
[\pi^\m,\pi ^\n]=-i e F^{\m\n}.
\label{GT}
\ceq 
In effect, the gauged equation is no longer scalar but becomes 
$(1/2,0)\oplus (0,1/2)$ representation specific, i.e. of four spinorial  dimensions.

In the following, when considering the gauging procedure, preference to the 
equations (\ref{newtree}) and  (\ref{newfour})    over  (\ref{j0cfact}) and (\ref{eco}) will be given. 
{}For free particles instead, the equations  
(\ref{dir_12}) and (\ref{ceq1}) will be systematically 
used.  The free $so(1,3)$ irreducible spin-$1/2^+$ and spin-$1/2^-$ degrees of freedom in the four-vector-- spinor, together with the
respective free particle wave equations, are listed in the Table 2. 

\begin{table} 
\begin{tabular}{|c||c|} \hline 
 {$so(1,3)$ irreducible free spin-$1/2$ degrees of freedom}& Free particle wave equations   \\ 
\hline 
~&~\\
\underline{spin-$1/2^+\in (1/2,0)\oplus (0,1/2)$}:  &  
\underline{In the full space:}\\
~&~\\
~&$\(\frac{1}{4}\left( \g_\a \slashed{p}\slashed{p}\g_\b\right)_{ab} -m^2\delta_{ab} g_{\a\b}\)\left[w^{(1/2, 
0)}_\pm\left({\mathbf p},\frac{1}{2},\lambda\right)\right]^{\b b}=0$ \quad
eq.~(\ref{dir_12})    \\ 
&~\\ 
$\[w_\pm^{(1/2,0)}\left({\mathbf p},\frac{1}{2},\lambda 
\right)\]^{\a}$=$\frac{1}{2 m}\g^\a \slashed{p} 
u_{\pm}\left(\mathbf{p},\lambda\right)$&    \underline{Contracted:} \\
~&~\\
~&      $\(p^2 -i\frac{g_{(1/2^+)}}{4}\sigma^{\mu\nu}\left[p_\mu,p_\nu\right]-m^2\)\left[ \gamma \cdot 
w^{(1/2,0)}_\pm\left({\mathbf p}, \frac{1}{2}, \l\right)\right] = 
0$\\
~&~\\
~&\quad ~eq.~(\ref{newtree})    ~\\
~&~\\
~&~\\ 
~&~\\
\hline 
~&~\\
\underline{spin-$1/2^- \in (1/2,1)\oplus (1,1/2)$}:&\underline{In the full space:}\\
~&~\\
~&~\\
$\[w_\pm^{(1/2,1)}\left({\mathbf p},\frac{1}{2},\lambda 
\right)\]^\a$=$\frac{2}{\sqrt{3} m}(p^\a -\frac{1}{4} 
\g^\a\slashed{p})u_{\pm}\left(\mathbf{p},\lambda \right)$&$\[\frac{4}{3}\(p_\a\delta_{ab}-\frac{1}{4}\left[\g_\a\right]_{ab}\slashed{p}\)\(p_\b\delta_{ab}-\frac{1} {4}\slashed{p}\left[\g_\b\right]_{ab}\)
-m^2 \delta_{ab}g_{\a\b}\]$\\
~&~\\
~&$\times \left[ w^{(1/2, 1)}_\mp \left({\mathbf p},\frac{1} {2},\l\right)\right]^{\b b}=0$ \quad eq.~(\ref{ceq1}) \\
~&~\\
~&\underline{Contracted:}\\
~&~\\ 
& 
$\left( p^2-i\frac{g_{(1/2^-)}}{4}\sigma^{\mu\nu}\left[p_\mu,p_\nu \right] -m^2\right)
\left[ p\cdot w^{(1/2, 1)}_\pm \left({\mathbf p},\frac{1}{2},\l\right)\right]=0$ \\
~&~\\
 ~&\,\, eq.~(\ref{newfour})\\
~&~\\
\hline
\end{tabular} 
\caption{ \label{table}
Summary of the  $so(1,3)$ irreducible degrees of freedom spanning the four-vector--spinor space (left column), 
and their description by second order wave equations (right column) 
free from auxiliary conditions. We list for each one of the spin-$1/2^+\in (1/2,0)\oplus (0,1/2)$, and spin-$1/2^-\in (1/2,1)\oplus (1,1/2)$ 
sectors two types of wave equations, 
the first referring to  the full  16 dimensional four-vector--spinor space, and the second, obtained from the previous one by either $\gamma^\a$, or, $p^\a$, contraction,
referring to four spinorial degrees of freedom. In section 5 below we write down the Lagrangiangs associated with all these equations, their gauged forms being  given then in Section 6.  In section 6 we show that Compton scattering off spin-$1/2^-$ does not distinguish between the full-space-- and the contracted Lagrangians. From that we conclude that the covariant spin-$1/2^-$--spin-$3/2^-$ separation performed at the free particle level is respected by the gauging procedure and that our approach continues describing the gauged four spin-$1/2^-$ by spinorial  degrees of freedom, as it should be.
On this basis we conclude that the spin-$1/2^-$ particle under discussion effectively behaves as a true ``quadratic'' fermion, as the contracted gauged quadratic equation
does not allow for a bi-linearization because of $g_{(1/2^-)}\not=2$. 
} 
\end{table}

{}Finally, the propagators of the two spin-$1/2$ sectors transforming according to different 
$j_1$-labels are obtained as the inverse to 
the respective equation operators as, 
\aeq\label{props0} 
~[S^{(1/2, j_1)}(p)]_{\a\b}=\([\G^{(1/2, j_1)}_{\m\n}]_{\a\b}p^\m p^\n-m^2 
g_{\a\b}\)^{-1}, 
\ceq 
and are  given by 
\aeq\label{props} 
~[S^{(1/2, j_1)}(p)]_{\a\b}=\frac{[\Delta^{(1/2, j_1)}(p)]_{\a\b}}{p^2- 
m^2+i\e}, 
\ceq 
where 
\aec 
~[\D^{(1/2, 0)}(p)]_{\a\b}&=&\frac{1}{m^2}\(\frac{1}{4}p^2\g_\a \g_\b+(m^2- 
p^2)g_{\a\b}\),\label{deltaj0}\\ 
~[\D^{(1/2, 1)}(p)]_{\a\b}&=&\frac{1}{m^2}\[\frac{4}{3}\(p_\a-\frac{1} 
{4}\g_\a\slashed{p}\)\(p_\b-\frac{1}{4}\slashed{p}\g_\b\)+(m^2- 
p^2)g_{\a\b}\]\label{deltaj1}. 
\cec 
In terms of the  Lorentz- and Poincar\'e- projectors from the respective eqs.~(\ref{pf0})-(\ref{pf1}), and (\ref{pro_12}),
the latter quantities are equivalent to, 
\aec 
~[\D^{(1/2, j_1)}(p)]_{\a\b}&=&\frac{p^2}{m^2}[\mathcal{P}^{(1/2,j_1)}_F]_{\a} 
{}^\g[\mathbb{P}^{(1/2)}(p)]_{\g\b}+\frac{(m^2-p^2)}{m^2}g_{\a\b}, 
\cec 
while in terms of the $[f^{(1/2, j_1)}(p)]^\a$ matrices one has, 
\aec 
~[\D^{(1/2, j_1)}(p)]_{\a\b}&=&[f^{(1/2, j_1)}(p)]_\a [\overline{f}^{(1/2, 
j_1)}(p)]_\b+\frac{(m^2-p^2)}{m^2}g_{\a\b}. 
\cec

\begin{quote} 
The conclusion is that  while the 16 dimensional $so(1,3)$ irreducible four-vector spinors describing its spin-$1/2^+$ and spin-$1/2^-$
sectors do not satisfy the Dirac equation, their contractions by $\gamma^\a$ and $p^\a$ down to four spinorial degrees of freedom each,
 do. Within this context the question arises, as to what extend the contracted four-vector spinors are eligible for the
description of the two spin-$1/2^+$ and spin-$1/2^-$ particles under discussion.
{}For the time being, and at  the free particle level, no distinction can be made 
between these  degrees of freedom, namely, between $\left[ \gamma\cdot w^{(1/2,0)}_\pm\left({\mathbf p}, \frac{1}{2}, \l\right)\right]$, 
and $\left[ p \cdot w^{(1/2,1)}_\mp\left({\mathbf p}, \frac{1}{2}, \l\right)\right]$,
in the respective equations (\ref{glory}) and (\ref{dirgdw2}),
because they both satisfy the free Dirac equation, and behave as spin-$1/2^+$, and spin-$1/2^-$, respectively. 
This situation is to change upon introducing interactions, an issue handled in 
the subsequent section. 
\end{quote}

 
 
 

\section{The electromagnetically  gauged contracted  wave equations }\label{sec4} 

 
 

 

In the current and the subsequent  sections we consider the couplings of the free spin-$1/2^+$ and spin-$1/2^-$ particle equations,
(\ref{dir_12}), (\ref{newtree}), and (\ref{ceq1}), (\ref{newfour}) to the electromagnetic field,
confining  to minimal gauging in the sense that  the coupling is defined through  changing 
ordinary by covariant derivatives according to, 
\aeq 
\partial^\mu\longrightarrow  
D^\mu=\pd^\mu+i e A^\mu , 
\label{gauge_simtrnsf} 
\ceq 
where  $e$ is the electric charge of the particle. 
In order to obtain the gauged equations in the full space, we first write \eqref{eomsj1} 
in position space for a plane wave of the type $[\y_\pm^{(1/2,j_1)}]^\a(x)\longrightarrow
[w^{(1/2, j_1)}_\pm({\mathbf p}, 1/2,\l )  ]^\a e^{\mp i\,x\cdot p}$,
 as 
\aeq 
\([{\widetilde \G}^{(1/2, j_1)}_{\m\n}]_{\a\b}\pd^\m \pd^\n+m^2 
g_{\a\b}\)[\y^{(1/2,j_1)(x)}_\pm(x)]^\b=0.\label{fdEqs} 
\ceq 
This  is in reality a  $(16\times 16)$ dimensional matrix equation for 
the $16$-component vector-spinor $[\y^{(1/2,j_1)}_\pm(x)]^\b$. 
However,  we have shown in the momentum space  equations ~(\ref{j0cfact}) and 
(\ref{dirgdw1}}) above 
that for spin-$1/2$ in $\psi_\mu$ 
the contractions of the latter  be it by  $\gamma^\alpha$, or $p^\alpha$,  are  four-dimensional spinors, 
which translates to position space as,
$\left[ \gamma\cdot \y^{(1/2,0)}_\pm(x)\right]$, and $\left[i\partial \cdot \y^{(1/2,0)}_\pm(x)\right]$, respectively. 
Below we gauge these two schemes and reveal their equivalence.

 
\subsubsection{Gauging the  contracted  wave  equation for the $so(1,3)$ irreducible \texorpdfstring{$(1/2,0)\oplus(0,1/2)$}{} sector}\label{sec4a1} 
 
Replacing now in (\ref{newtree}) everywhere  the ordinary by the covariant derivatives, 
$\partial^\mu\longrightarrow D^\mu$, gives 

\aeq 
\({D}^2+g_{(1/2^+)}\frac{e}{4}\sigma^{\mu\nu}F_{\mu\nu} +m^2\)
\left[ \gamma\cdot \y_\pm^{(1/2,0)}(x)\right]=0.  
\label{Geq0} 
\ceq 
{}For $g_{(1/2^+)}=2$ the gauged $\left[ \gamma\cdot \y_\pm^{(1/2,0)}(x)\right]$ spin-$1/2^+$ solution in $(1/2,0)\oplus (0,1/2)$ ,
would be proportional to the gauged Dirac spinor,
\begin{equation}
 \slashed{D}\left[ \gamma\cdot \y_\pm^{(1/2,0)}(x)\right]=-im\left[ \gamma\cdot  \y_\pm^{(1/2,0)}(x)\right] \Longrightarrow
 \left[\gamma\cdot \y_\pm^{(1/2,0)}(x)\right] \sim { \y}_\pm^{D}(x).
\label{result1}  
\end{equation}
In other words, both   $\left[\gamma\cdot \y_\pm^{(1/2,0)}(x)\right]$, and
 ${ \y}^D_\pm(x)$ now solve the gauged Dirac equation \cite{Mikheev} and describe same spin-$1/2^+$  particle of a
gyromagnetic factor $g_{(1/2^+)}=2$.

 
\subsubsection{Gauging the contracted  wave equation for spin-\texorpdfstring{$1/2^-$}{}  in the $so(1,3)$  irreducible 
 \texorpdfstring{{$(1/2,1)\oplus(1,1/2)$}}{} sector}\label{sec4a2} 
 
Along same line as in the previous subsection gauging (\ref{newfour}) amounts to,
 \aeq 
\(D^2+ \frac{g_{(1/2^-)}}{4}e\sigma^{\mu\nu}F_{\mu\nu}  +m^2\)  
\[D\cdot  \psi_\mp^{(1/2,1)}(x)\]=0, 
\label{new_spnr} 
\ceq 
where ${ g} _{(1/2^-)}=-2/3$ takes the place of the physical gyromagnetic factor of the spin-$1/2^-$ particle under consideration.
Remarkably, both the equations (\ref{Geq0}) and (\ref{new_spnr}) are of the art of the equation (\ref{KGG1}) following from the Lagrangian
in (\ref{KGG2}).
This circumstance will allow us to decide whether 
$\[D\cdot  \psi_\mp^{(1/2,1)}(x)\]$ behaves as a Dirac-, or, as a quadratic fermion,
in depending on whether or not its $ { g} _{(1/2^-)}$ value  happens to be confirmed  by the magnetic dipole moment calculated from the
electromagnetic current in the full 16 dimensional four-vector spinor space.
In the next section we shall present the Lagrangians associated with the gauged equations within the full  space,
obtain the electromagnetic currents, and calculate the magnetic dipole moments for both the spin-$1/2$ sectors under consideration.
This will allow us to figure out the true nature of spin-$1/2^-$.

 

\section{The gauged Lagrangians of  the two spin-\texorpdfstring{$1/2$}{} sectors of the four-vector field within the full space}\label{sec4b} 
 
\begin{figure} 
\includegraphics{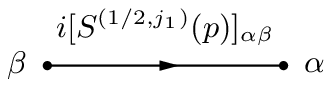} 
\caption{\label{propfig} Feynman rule for the propagators of particles in the 
$(1/2,j_1)\oplus(j_1,1/2)$ sector of the four-vector-spinor. They are obtained 
as the inverse of the equations of motion, the explicit form of $S^{(1/2,j_1)}
(p)$ 
for each $j_1$-value is given in \eqref{props0}-\eqref{deltaj1}.} 

\includegraphics{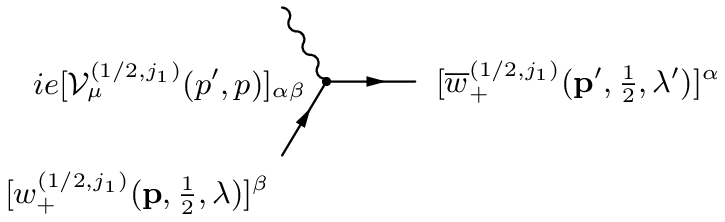} 
\caption{\label{regla1fig} Feynman rule for the one-photon vertex with 
$(1/2,j_1)\oplus(j_1,1/2)$ particles. It comes from the interaction Lagrangian 
\eqref{lint}, for the specific definition of 
$\mathcal{V}^{(1/2,j_1)}_\m(p',p)$ 
corresponding to each $j_1$-value see \eqref{vmdef}.} 

\includegraphics{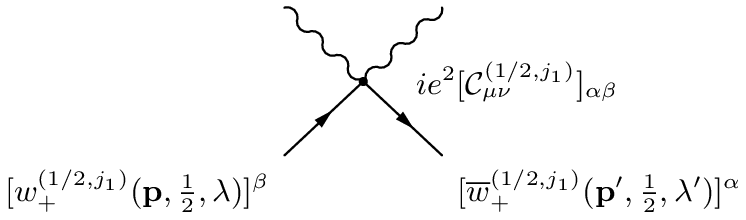} 
\caption{\label{regla2fig} Diagram for the two-photons Feynman rule 
corresponding to the interaction Lagrangian \eqref{lint}, the explicit form of 
the $\mathcal{C}^{(1/2,j_1)}_{\m\n}$ vertex is given in \eqref{cmndef}.} 
\end{figure}

{}For practical calculations it is advantageous  to have at our disposal  
gauged Lagrangians, 
from which one can extract  the Feynman rules of the theory. The Lagrangians 
for positive and negative parity states differ only by an overall sign that  
reflects  the relative sign 
in the  normalizations of the opposite parity states. 
In the following we shall only deal with Lagrangians written in terms of the 
positive parity states. 
The free equations of motion 
\eqref{dir_12} and (\ref{eco}) relate 
to  Lagrangians in the standard way as  Euler-Lagrange equations. For second order equations 
of motion the corresponding Lagrangians 
are of the form, 
\aeq 
\mathcal{L}^{(1/2,j_1)}_{\text{free}}(x)=(\pd^\m[\overline{\y}^{(1/2, 
j_1)}_+ (x)]^\a) 
[{\widetilde \G}^{(1/2, j_1)}_{\m\n}]_{\a\b} \pd^\n [\y_+^{(1/2, j_1)}(x)]^\b 
-m^2[\overline{\y}_+^{(1/2, j_1)}(x)]^\a[\y_+^{(1/2, j_1)}(x)]_\a, 
\ceq 
and the gauged Lagrangians are obtained as usually by the replacement of the 
ordinary-  by covariant derivatives, 
\aeq 
\mathcal{L}^{(1/2, j_1)}(x)=(D^{\m*}[\overline{\y}_+^{(1/2, j_1)}(x)]^\a) 
[{\widetilde \G}^{(1/2, j_1)}_{\m\n}]_{\a\b} D^\n 
[\y_+^{(1/2, j_1)}(x)]^\b-m^2[\overline{\y}_+^{(1/2, j_1)}(x)]^\a[\y_+^{(1/2, 
j_1)}(x)]_\a. 
\ceq 
After the use of  \eqref{dir_12}, the Lagrangian with 
$j_1=0$, which corresponds to the sector $(1/2,0)\oplus(0,1/2)$, assumes  the 
following explicit form, 
\aeq 
\mathcal{L}^{(1/2, 0)}(x)=\frac{1}{4}(D^{\m*}[\overline{\y}_+^{(1/2, 0)}(x)]^\a) 
\g_\a\g_\m 
\g_\n 
\g_\b [\y_+^{(1/2,0)}(x)]^\b-m^2[\overline{\y}_+^{(1/2,0)}(x)]^\a D^\m [\y_+^{(1/2, 
0)}(x)]_\a. 
\label{GI}
\ceq 
while the $j_1=1$ Lagrangian, which is the one relevant for the 
$(1/2,1)\oplus(1,1/2)$ sector reads, 
\begin{eqnarray}
\mathcal{L}^{(1/2, 1)}(x)&=&\frac{4}{3}(D^{\m*}[\overline{\y}_+^{(1/2, 1)}(x)]^\a) 
\(g_{\a\m}-\frac{1}{4}\g_\a\g_\m\)\(g_{\n\b}-\frac{1}{4}\g_\n \g_\b\)[\y_+^{(1/2,1)}(x)]^\b\nonumber\\
&-&m^2[\overline{\y}_+^{(1/2,1)}(x)]^\a[\y_+^{(1/2, 1)}(x)]_\a,
\label{GII}
\end{eqnarray}
where we made use of  \eqref{ceq1}. Now  $\mathcal{L}^{(1/2,j_1)}(x)$  can be 
decomposed  into free and interaction 
Lagrangians as, 
\aec 
\mathcal{L}^{(1/2, 
j_1)}(x)&=&\mathcal{L}^{(1/2,j_1)}_{\text{free}}(x)+\mathcal{L}^{(1/2,j_1)}_{\text{int}}(x),\\ 
\mathcal{L}^{(1/2,j_1)}_{\text{int}}(x)&=&-j_\m^{(1/2,j_1)}A^\m (x)+k^{(1/2, 
j_1)}_{\m\n}(x) A^\m (x) 
A^\n(x)\label{lint}. 
\cec 
Here, $j_\m^{(1/2,j_1)}(x)$ is the electromagnetic current density, while 
$k^{(1/2,j_1)}_{\m\n}(x)$ 
defines the structure of the two-photon coupling. 
Back to momentum space, and for the 
positive parity states $[w^{(1/2, j_1)}_+({\mathbf p},1/2,\l )]^\b$ we 
find, 
\aec 
j_\m^{(1/2,j_1)}\left( {\mathbf p},{\mathbf p}^\prime,\l,\l^\prime\right) 
&=& e\left[\overline{w}_+^{(1/2,j_1)}\left(\mathbf{p}',\frac{1}{2}, \l^\prime 
\right)\right]^\a 
[\mathcal{V}_{\m}^{(1/2,j_1)}( p',p)]_{\a\b} 
\left[w_+^{(1/2,j_1)}\left(\mathbf{p},\frac{1}{2},\l 
\right)\right]^\b,\label{currentj1}\\ 
k_{\m\nu}^{(1/2, j_1)}({\mathbf p},{\mathbf p}^\prime,\lambda,\lambda^\prime 
)&=&e^2\left[\overline{w}_+^{(1/2, j_1)}\left(\mathbf{p}',\frac{1}
{2},\l^\prime 
\right)\right]^\a[\mathcal{C}_{\m\n}^{(1/2, j_1)}]_{\a\b}\left[w_+^{(1/2, 
j_1)}\left( 
\mathbf{p},\frac{1}{2},\l \right)\right]^\b, 
\cec 
with 
\aec 
~[\mathcal{V}_{\m}^{(1/2, j_1)}(p',p)]_{\a\b}&=&
[{\widetilde \G}^{(1/2,j_1)}_{\n\m}]_{\a\b}p'^{\n}+ 
[{\widetilde \G}^{(1/2,j_1)}_{\m\n}]_{\a\b}p^{\n},\label{vmdef}\\ 
~[\mathcal{C}_{\m\n}^{(1/2,j_1)}]_{\a\b}&=&\frac{1}{2}\( 
[{\widetilde \G}^{(1/2,j_1)}_{\m\n}]_{\a\b}+[{\widetilde \G}^{(1/2,j_1)}_{\n\m}]_{\a\b}\)\label{cmndef}. 
\cec 
Here, $[\mathcal{V}_{\m}^{(1/2,j_1)}(p',p)]_{\a\b}$ and 
$[\mathcal{C}_{\m\n}^{(1/2, j_1)}]_{\a\b}$ are the respective one- and two-photon 
vertexes which, together with the propagators \eqref{props}, determine the 
Feynman rules. The latter are  depicted on the Figs. 
\ref{propfig},\ref{regla1fig},\ref{regla2fig}. 

In particular, $[\mathcal{V}_{\m}^{(1/2,j_1)}(p',p)]_{\a\b}$ obeys the 
Ward-Takahashi identity, 
\aeq 
(p'-p)^\m 
[\mathcal{V}_{\m}^{(1/2, j_1)}(p',p)]_{\a\b}=[S^{(1/2,j_1)}(p')]_{\a\b}^{-1}-
[S^{(1/2,j_1)} 
(p)]_{\a\b}^{-1}\label{wti}, 
\ceq 
where $[S^{(1/2,j_1)}(p)]_{\a\b}$ are the propagators in \eqref{props}. 
This relationship  implies gauge invariance of the 
amplitudes which define the Compton scattering process. 
However, before evaluating  this very 
process it is instructive to calculate 
the values of the magnetic dipole moments of the particles under 
consideration, as they appear 
prescribed by the currents in \eqref{currentj1}. 
 
 
\subsection{Magnetic dipole moments}\label{sec4c} 
 
We now rewrite  the momentum space currents in 
\eqref{currentj1} in terms of free Dirac $u$-spinors as,
\aeq\label{j1ov} 
j_\m^{(1/2,j_1)}(\mathbf{p},\mathbf{p'},\l ,\l^\prime)= e \overline{u}_+ 
(\mathbf{p}',\l^\prime ) 
\widetilde{\mathcal{V}}^{(1/2, j_1)}_\m({p}',{p}) u_+(\mathbf{p},\l). 
\ceq 
Here we have used the equations (\ref{defl0s})-(\ref{defl1s})  to come to, 
\aeq 
~\left[w_\pm^{(1/2, j_1)}\left(\mathbf{p},\frac{1}{2},\l\right)\right]^\a= 
[f^{(1/2, j_1)}(p)]^\a 
u_{\pm}(\mathbf{p},\l)\label{deflf},
\ceq 
and on the basis of this relationship defined the new vertex, $\widetilde{\mathcal{V}}^{(1/2, j_1)}_\m({p}',{p})$, as 
\aec 
\widetilde{\mathcal{V}}^{(1/2, j_1)}_\m({p'},{p})&=&[\overline{f}^{(1/2, j_1)}
({p}')]^\a 
[\mathcal{V}_{\m}^{(1/2,j_1)}(p',p)]_{\a\b}[f^{(1/2, j_1)}({p})]^\b, \quad j_1=0,1.
\cec 
Incorporation of the  mass-shell condition, and $\slashed{p}\, u_+({\mathbf p},\lambda)=mu_+({\mathbf p},\lambda)$, 
amounts to, 
\aec 
j_\m^{(1/2, 0)}(\mathbf{p'},\mathbf{p},\l,\l^\prime )&=& e\, \overline{u}_+ 
(\mathbf{p}',\l^\prime )\( 
2m\g_\m \)u_+(\mathbf{p},\l),\label{j0gd}\\ 
j_\m^{(1/2, 1)}(\mathbf{p'},\mathbf{p},\l,\l^\prime )&=& e\, \overline{u}_+ 
(\mathbf{p}',\l^\prime ) 
\(\frac{4}{3}(p'+p)_\m-\frac{2m}{3}\g_\m \) u_+(\mathbf{p},\l)\label{j1gd}. 
\cec 
We now perform  the Gordon decomposition of the first  current, yielding, 
\aeq 
2m\,e\, \overline{u}_+(\mathbf{p}',\l^\prime )\g_\m 
u_+(\mathbf{p},\l)=e\,\overline{u}_+(\mathbf{p}',\l^\prime )\[ (p'+p)_\m+2 i 
M^S_{\m\n}(p'-p)^\n \]u_+(\mathbf{p},\l). 
\ceq 
Here $M^S_{\m\n}$ are the elements of the Lorentz algebra in $(1/2,0)\oplus 
(0,1/2)$ in \eqref{genss}, while the factor $2$ in front of them stands for 
the gyromagnetic ratio. Therefore,
one immediately notices that for $j_1=0$, the textbook Dirac 
current with $g_{(1/2^+)}=2$ is recovered. 
Similarly, the second current is handled.
As a result, both  currents in \eqref{j0gd} and \eqref{j1gd} 
are equally shaped after, 
\begin{eqnarray} 
j_\m^{(1/2,j_1)}(\mathbf{p'},\mathbf{p},\l,\l^\prime )&=&e\,\overline{u}_+ 
(\mathbf{p}',\l^\prime )\[ 
(p'+p)_\m+i g_{s_i}M^S_{\m\n}(p'-p)^\n \]u_+(\mathbf{p},\l),\nonumber\\
s_1&=&1/2^+,\quad s_2= 1/2^-,
\end{eqnarray}
with
\aec 
g_{(1/2^+)}&=&2\label{g0},\\ 
g_{(1/2^-)}&=&-\frac{2}{3}\label{g1}. 
\cec 
The equations ~\eqref{g0} and \eqref{g1} show that the electromagnetic 
currents for particles transforming in 
$(1/2,j_1)\oplus(j_1,1/2)$ are characterized by different magnetic dipole 
moments for the two
different $j_1=0,1$ values in $(1/2,j_1)\oplus (j_1,1/2)$. The gauged 
Lagrangian corresponding to the combined  Lorentz- and Poincar\'e invariant 
projector, 
that describes particles of charge $e$  transforming in 
$(1/2,0)\oplus(0,1/2)$ predicts the following magnetic moment, 
\aeq 
\m_{(1/2^+)}(\l)=2\frac{\l e}{2 m}\label{dipmm0}. 
\ceq 
The latter coincides with the standard value for a Dirac particle of 
polarization $\l$. 
Instead,  the Lagrangian of same type  predicts for  particles of charge $e$ 
transforming in  $(1/2,1)\oplus(1,1/2)$  a magnetic dipole moment of 
\aeq 
\m_{(1/2^-)}(\l)=-\frac{2}{3}\frac{\l e}{2 m}\label{dipmm1}. 
\ceq 
Although these $g_s$ values coincide with those in (\ref{newtree}) and (\ref{newfour}), it is precipitate to claim equivalence between the
effective spinorial degrees of freedom solving the contracted equations, and the genuine four-vector-spinor degrees of freedom describing the two spin-$1/2$ sectors under consideration. The reason is that an electromagnetic process is not entirely determined by the 
electromagnetic multipole moments of the particles, 
which by definition are associated with the on-shell states, 
it is determined  by the 
complete gauged Lagrangian. The mere 
confirmation  of the  electromagnetic multipole  moments 
by a theory is  not sufficient to claim its credibility. 
This because different Lagrangians can predict 
equal multipole moments.\cite{Lorce:2009bs}{},\cite{DelgadoAcosta:2012yc} {}.  
The more profound test for the predictive power of  Lagrangians concerns 
processes 
involving off-shell  states. One such process, the Compton scattering, 
is the subject of  the next section. 
 
 

\section{ Compton scattering off spin-\texorpdfstring{$1/2$}{} in 
\texorpdfstring{$(1/2,j_1)\oplus(j_1,1/2)$} {} with    \texorpdfstring{$j_1=0,1$}{}}\label{sec5} 

\subsection{The calculation within the full space}

\begin{figure} 
\includegraphics{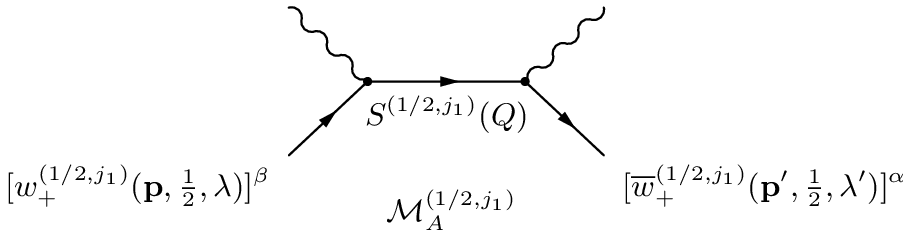} 
\caption{\label{mafig} Diagram for the direct-scattering contribution 
\eqref{maj1} to the Compton scattering amplitude \eqref{camp}.} 

\includegraphics{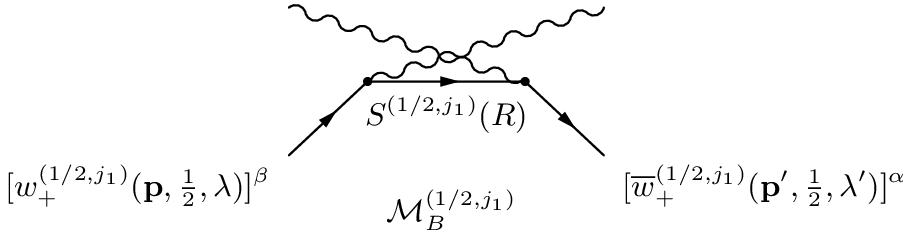} 
\caption{\label{mbfig} Diagram for the exchange-scattering contribution 
\eqref{mbj1} to the Compton scattering amplitude \eqref{camp}.} 

\includegraphics{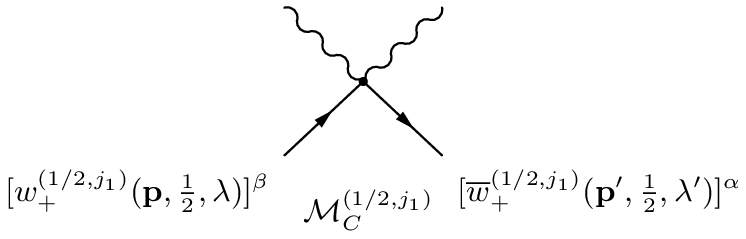} 
\caption{\label{mcfig} Diagram for the point-scattering contribution 
\eqref{mcj1} to the Compton scattering amplitude \eqref{camp}.} 
\end{figure}

\noindent
The construction of the Compton scattering amplitudes from the Feynman 
rules (shown in the Figs. \ref{propfig},\ref{regla1fig}, and \ref{regla2fig}) 
for 
each $j_1$- value, is standard \cite{Bjorken} and reads, 
\aeq\label{camp} 
\mathcal{M}^{(1/2, 
j_1)}=\mathcal{M}_A^{(1/2,j_1)}+\mathcal{M}_B^{(1/2,j_1)}+\mathcal{M}_C^ 
{(1/2,j_1)}, 
\ceq 
where $\mathcal{M}_A^{(1/2,j_1)}$, $\mathcal{M}_B^{(1/2, j_1)}$, 
$\mathcal{M}_C^{(1/2, j_1)}$ correspond to the amplitudes 
for direct, exchange and contact  scatterings, respectively. 
In the following we 
use  $p$ and $p'$ to denote the momentum of the incident and 
scattered spin-$1/2$ particles respectively, while $q$ and $q'$ stand in their 
turn for 
the momenta of the incident and scattered photons. 
In effect, we find, 
\aec 
i\mathcal{M}_A^{(1/2, j_1)}= 
e^2 
\left[ \overline{w}_+^{(1/2, j_1)} 
\left( 
\mathbf{p}',\frac{1}{2},\l' 
\right)\right]^{\a} 
~[U^{(1/2,j_1)}_{\m\n}(p',Q,p)]_{\a\b} 
&&\left[ 
w_+^{(1/2, j_1)} 
\left( 
\mathbf{p},\frac{1}{2},\l 
\right)\right]^{\b} \nonumber\\
&\times &[\e^\m(\mathbf{q}',\ell')]^* 
\e^\n(\mathbf{q},\ell), 
\label{maj1}\\ 
i\mathcal{M}_B^{(1/2,j_1)}=e^2\left[\overline{w}_+^{(1/2,j_1)}\left(\mathbf{
p}',\frac{ 
1}{2},\l'\right)\right]^{\a}~[U^{(1/2,j_1)}_{\n\m}(p',R,p)]_{\a\b} 
&&\left[ w_+^{(1/2,j_1)}\left(\mathbf{p},\frac{1}{2},\l\right)\right]^{\b}\nonumber\\
&\times & [\e^\m(\mathbf{q}',\ell')]^* 
\e^\n(\mathbf{q},\ell),\label{mbj1}\\ 
i\mathcal{M}_C^{(1/2,j_1)}=-e^2\left[\overline{w}_+^{(1/2,j_1)}\left( 
\mathbf{p}',\frac{1}{2},\l' \right)\right]^{\a}~ 
[\mathcal{C}_{\m\n}^{(1/2,0)}+\mathcal{C}_{\n\m}^{(1/2,0)}]_{\a\b} 
&&\left[ w_+^{(1/2, j_1)}\left(\mathbf{p},\frac{1}{2},\l\right)\right]^{\b}\nonumber\\ 
&\times &[\e^\m(\mathbf{q}',\ell')]^* 
\e^\n(\mathbf{q},\ell)\label{mcj1}, 
\cec 
where $Q=p+q=p^\prime +q'$ and $R=p'-q=p-q'$ stand for the momentum of the 
intermediate states, while  
\aeq 
~[U^{(1/2,j_1)}_{\m\n}(p',Q,p)]_{\a\b}=[\mathcal{V}^{(1/2,j_1)}_{\m}
(p',Q)]_{\a\g} 
[S^{(1/2,j_1)}(Q)]^{\g\d}[\mathcal{V}^{(1/2,j_1)}_\n(Q,p)]_{\d\b}. 
\ceq 
These amplitudes are shown in the Figs. \ref{mafig},\ref{mbfig},\ref{mcfig}. 
Their gauge invariance  is ensured by the 
Ward-Takahashi identity \eqref{wti} (c.f. 
\cite{DelgadoAcosta:2009ic}). 

It is furthermore quite useful to define the following quantities: 
\aec 
\widetilde{U}^{(1/2, j_1)}_{\m\n}(p',Q,p)&=&[\overline{f}^{(1/2,j_1)}
({p}')]^\a 
[U^{(1/2,j_1)}_{\m\n}(p',Q,p)]_{\a\b}[f^{(1/2,j_1)}({p})]^\b,\label{utrans}\\ 
\widetilde{\mathcal{C}}^{(1/2,j_1)}_{\m\n}&=&[\overline{f}^{(1/2,j_1)}
({p}')]^\a 
[\mathcal{C}_{\m\n}^{(1/2,j_1)}+\mathcal{C}^{(1/2, j_1)}_{\n\m}]_{\a\b} 
[f^{(1/2, j_1)}({p})]^\b,\label{ctrans} 
\cec 
with the $[f^{(1/2,j_1)}({p})]^\a$ matrices taken from \eqref{ef0} and 
\eqref{ef1}.  

In order to find the averaged square amplitude,  
we employ the following general expression: 
\aec\label{m2av} 
\overline{\left\vert 
\mathcal{M}^{(1/2,j_1)}\right\vert^2}&=&\frac{1}{4}\sum_{\l,\l',\ell,\ell'} 
[\mathcal{M}^{(1/2,j_1)}][\mathcal{M}^{(1/2,j_1)}]^\dagger\\ 
&=&Tr\left[(\mathcal{M}^{(1/2,j_1)})_{\m\n}(p',Q,R,p) 
(\mathcal{M}^{(1/2,j_1)})^{\n\m}(p,R,Q,p')\right], 
\cec 
valid for any $j_1$. Here,  we have defined 
\aec 
\mathcal{M}^{(1/2,j_1)}_{\m\n}(p',Q,R,p)&=&\frac{e^2}{2}\
\left(\frac{\slashed{p}'+m} 
{2m}\right)\mathbb{U}^{(1/2, j_1)}_{\m\n}(p',Q,R,p),\\ 
\mathbb{U}^{(1/2,j_1)}_{\m\n}(p',Q,R,p)&=&\widetilde{U}^{(1/2,j_1)}_{\m\n} 
(p',Q,p)+\widetilde{U}^{(1/2,j_1)}_{\n\m}(p',R,p)- 
\widetilde{\mathcal{C}}^{(1/2, j_1)}_{\m\n}.
\cec 
{}Furthermore,  $\widetilde{U}$ and $\widetilde{\mathcal{C}}$ have been defined  in 
\eqref{utrans}, \eqref{ctrans}, and use has been made of the projector, 
\aeq 
~[\mathbb{P}^{(1/2,j_1)}_+({p})]_{\a\b}=[f^{(1/2, j_1)}({p})]_\a\
\left(\frac{\slashed{p}+m} {2m}\right)
[\overline{f}^{(1/2, j_1)}({p})]_\b,
\ceq 
in combination with the projector on the photon polarization vectors, 
\aeq 
\sum_{\ell}\e^\m(\mathbf{q},\ell)[\e^\n(\mathbf{q},\ell)]^{*}=- 
g^{\m\n}.\label{phpr}
\ceq 

The result for each $j_1$ value can be expressed by a single  general formula valid 
for any $g_{s_i}$, as: 
\begin{eqnarray}\label{m2nkr12j1} 
\overline{\vert{\mathcal M}^{(1/2,j_1)}(g^{(1/2,j_1)})\vert^2}&=& f_0+f_D 
+\frac{e^4(2 m^2-s-u)}{16 m^2 \left(m^2-s\right)^2 
\left(m^2-u\right)^2}\sum_{k=1}^4 (g_{(s_i)}-2)^k a_{k}, 
\end{eqnarray} 
where: 
\begin{eqnarray} 
a_1&=&-32 m^2 \left(m^2-s\right) \left(m^2-u\right) \left(2 
m^2-s-u\right),\label{ca1}\\ 
a_2&=&-4 \left(13 m^8-17 (s+u) m^6+\left(6(s^2+u^2)+20 u s\right) m^4-7 s u 
(s+u) m^2+3 s^2 u^2\right),\\ 
a_3&=&-8 \left(m^2-s\right)^2 \left(m^2-u\right)^2,\\ 
a_4&=&\left(m^2-s\right) \left(m^2-u\right) \left(m^2 (s+u)-2 s 
u\right),\label{ca4} 
\label{Gl2} 
\end{eqnarray} 
and with  $f_0$, $f_D$ standing for, 
\aec\label{def:f0} 
f_0&=& \frac{4e^4(5m^8-4(s+u)m^6+(s^2+u^2)m^4+s^2 u^2)}{(m^2-s)^2(m^2-u)^2},\\ 
f_D&=&-\frac{2e^4(-2m^2+s+u)^2}{(m^2-s)(m^2-u)}\label{def:fd}. 
\cec 
Here, $s$ and $u$, are the standard Mandelstam variables.

A comment is due  on the unspecified parity of the  propagator, typical for 
all second order approaches. 
We here gain control over the issue through the systematic explicit 
specification of the parities of  the external legs. 
Notice that none of the Feynman diagrams entering the 
calculation of the cross sections contains 
opposite parity states as external legs. At most, such a state could  appear 
in the internal lines. However, 
as long as such diagrams would be equivalent to diagrams with  a chirality operator 
inserted in each one of the vertexes, they will be  indistinguishable from the regular diagrams with an internal line consistent 
with the initial state, and can be  accounted for  by a proper normalization.
 Moreover, we investigated whether Feynman diagrams with legs of opposite 
parities, i.e. such invoking an axial electromagnetic current in one of the vertexes,  could 
contribute to the Compton scattering process under investigation and 
found them  vanishing within the $(1/2,0)\oplus (0,1/2)$ sector for $g_s=2$. In this way, the parity conservation 
is strictly respected  by the spin-$1/2^+$ Dirac particle, as it should be.
The spin-$1/2^-$,  and the (here omitted) spin-$3/2^-$ residents in $(1/2,1)\oplus (1,1/2)$, however, are different and such diagrams are non vanishing for all $g_s$ values, certainly an interesting issue,  worth being pursued in future research.
{}For the time being,  we limit ourselves to the observation that such fermions may be eligible as matter fields in dual theories of electromagnetism with 
co-existing axial  and regular photons. 

Obtaining now the differential cross-section in the laboratory frame from the 
squared amplitudes  in 
\eqref{m2nkr12j1} is  straightforward (see \cite{DelgadoAcosta:2010nx} for 
technical details). 
After some algebraic manipulations one arrives at, 
\aeq\label{dss} 
\frac{d\s(g_{(s_i)},\h,x)}{d \Omega}= 
\frac{\overline{\left\vert \mathcal{M}^{(1/2,j_1)}
(g_{(s_i)})\right\vert^2}}{4(4\pi m^2)}\frac{\left(\omega^\prime\right)^2 
}{\omega^2} 
= 
z_0+z_D+\frac{(x-1) r_0^2}{64 ((x-1) 
\eta 
-1)^3}\sum_{k=1}^4 (g_{(s_i)}-2)^k b_k, 
\ceq 
where $r_0=e^2/(4\p m)=\a m$, $\h=\o/m$ with $\o$  being the energy if the 
incident 
photon, $\omega^\prime $ the energy of the scattered photon, while $x=\cos\q$, 
$\q$ giving the scattering angle in the laboratory 
frame. 
In \eqref{dss} $z_0$ denotes the standard differential cross-section for 
Compton scattering 
off spin-0 particles and $(z_0+z_D)$  is the standard differential 
cross-section for 
Compton scattering off Dirac particles, i.e., 
\begin{eqnarray} 
z_0&=&\frac{\left(x^2+1\right) r_0^2}{2 ((x-1) \eta -1)^2},\label{z0}\\ 
z_D&=&-\frac{(x-1)^2 \eta ^2 r_0^2}{2 ((x-1) \eta -1)^3}. \label{zd} 
\end{eqnarray} 
We further have introduced the following notations, 
\begin{align} 
b_1=&-32 (x-1) \eta ^2,\\ 
b_2=&4 \left(x^2-3 x+8\right) \eta ^2,\\ 
b_3=&16 \eta ^2,\\ 
b_4=&(x+3) \eta ^2. 
\end{align} 

\begin{figure}[ht] 
\includegraphics{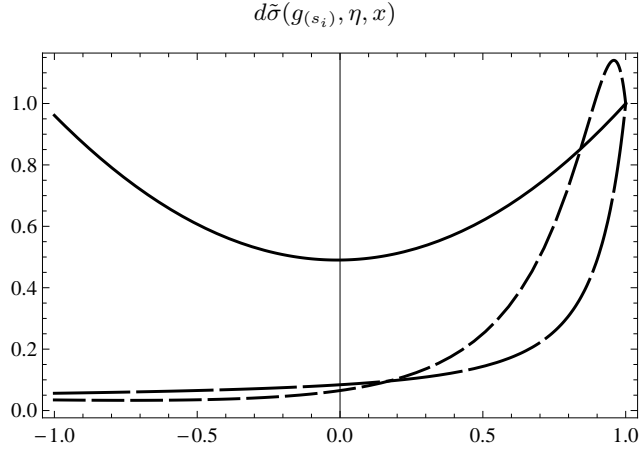} 
\caption{\label{dsfig} Differential cross sections for particles in the 
$(1/2,j_1)\oplus(j_1,1/2)$ sector of the four-vector spinor as a function of 
$x=\cos\q$  (where $\q$ is the scattering angle). The solid curve represents 
the classical limit, i.e. the differential cross section 
$d\widetilde{\s}(g_{(s_i)},\h,x)$ from \eqref{dst} at $\h=\omega/m=0$ 
(where 
$\omega$ is the energy of the incident photon), the long-dashed curve 
corresponds to $d\widetilde{\s}(g_{(1/2^+)},\h,x)$ with $g_{(1/2^+)}=2$ at the 
energy 
of 
$\h=4$, and the short-dashed curve represents $d\widetilde{\s}
(g_{(1/2^-)},\h,x)$ 
with $g_{(1/2^-)}=-2/3$ also at the energy of $\h=4$. The differential cross 
section 
has the correct Thompson limit for any $g_{(s_i)}$ value.} 
\end{figure}

The differential cross-section \eqref{dss} is found to have the following limits: 
\aec 
\lim_{x\rightarrow 1} \frac{d \s(g_{(s_i)},\h,x)}{d \Omega}&=&r_0^2,\\ 
\lim_{\h\rightarrow 0}\frac{d \s(g_{(s_i)},\h,x)}{d 
\Omega}&=&\frac{r_0^2}{2}(x^2+1),\\ 
\lim_{\h\rightarrow \infty}\frac{d \s(g_{(s_i)},\h,x)}{d \Omega}&=&0, 
\cec 
meaning that in the forward direction ($x=\cos\q =1$) it takes the  $r_0^2$ 
value. 
In the classical $\h \rightarrow 0$ limit the differential cross section is 
symmetric with respect to 
the scattering angle $\q$, while  in the high energy $\h\rightarrow 
\infty$ limit it vanishes independently of the $g_{(s_i)}$ factor value. 
This observation applies to each one of the two $j_1=0$-, and $j_1=1$ sectors 
of $\psi_\mu$ 
considered here,  and the related  $g_{(1/2^+)}=2$ and $g_{(1/2^-)}=-2/3$ 
values. 
The behavior of the differential 
cross-section is displayed  in  Fig. \ref{dsfig}, which is a plot of, 
\aeq 
d\widetilde{\s}(g_{(s_i)},\h,x)\equiv\frac{1}{r_0^2}\frac{d 
\s(g_{(s_i)},\h,x)}{d 
\Omega},
\label{dst}
\ceq 
for the two different $g_{(1/2^+)}=2$, $g_{(1/2^-)}=-2/3$ values of the gyromagnetic factors. 


Integration of \eqref{dss} over the solid angle leads to the total cross- 
sections, 
\aeq\label{tcs} 
\s(g_{(s_i)},\h)=s_0+s_D 
+\sum_{k=1}^4 (g^{(1/2,j_1)}-2)^k\(\frac{c_k}{128 \eta  (2 \eta +1)^2} 
+\frac{\log (2 \eta +1)h_k}{256 \eta ^2}\)3 \s_T, 
\ceq 
where $\s_T$ stands for the Thompson cross section $\s_T=(8/3)\pi r_0^2$. 
The following notations have been used, 
\aec 
s_0&=&\frac{3 (\eta +1) \s_T (2 \eta  (\eta +1) 
-(2 \eta +1) \log (2 \eta +1))}{4 \eta ^3 (2 \eta +1)},\label{s0}\\ 
s_D&=&\frac{3 \s_T \left((2 \eta +1)^2 \log (2 \eta +1) 
-2 \eta  (3 \eta +1)\right)}{8 \eta  (2 \eta +1)^2}\label{sd}, 
\cec 
where $s_0$ and $(s_0+s_D)$ are the standard cross-sections for Compton 
scattering off spin-$0$ and spin-$1/2$ Dirac particles, while 
the $c$ and $h$ coefficients stand for the following quantities, 
\aec 
c_1&=&-32 \eta  (3 \eta +1),\\ 
c_2&=&4 \left(6 \eta ^3+\eta ^2+8 \eta +3\right),\\ 
c_3&=&16 \eta ^3,\\ 
c_4&=&\eta  \left(4 \eta ^2+3 \eta +1\right),\\ 
h_1&=&32 \eta,\\ 
h_2&=&4 (\eta -3),\\ 
h_3&=&0,\\ 
h_4&=&-\eta. 
\cec 
The total cross section \eqref{tcs} has the following limits, 
\aec 
\lim_{\h\rightarrow 0}\s(g_{(s_i)},\h)&=&\s_T,
\label{Thompson}\\ 
\lim_{\h\rightarrow 
\infty}\s(g_{(s_i)},\h)&=&\frac{3}{128}
(g_{(s_i)}-2)^2((g_{(s_i)})^2+2)\s_T. 
\label{qdrtc_frm}
\cec 
\begin{figure}[ht] 
\includegraphics{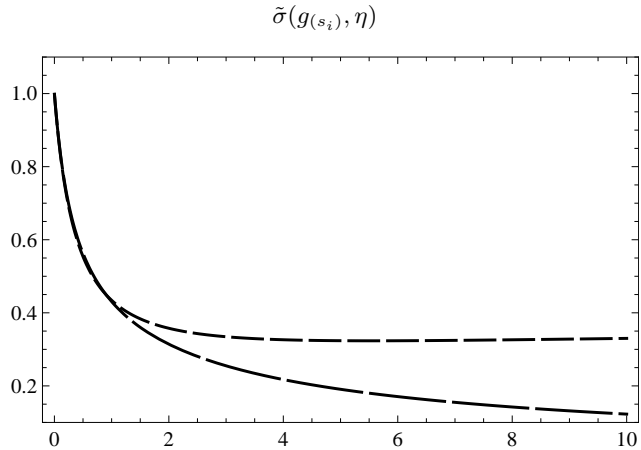} 
\caption{\label{sfig} The total cross sections $\widetilde{\s}(g_{(s_i)},\h)$ 
in eq.~\eqref{tcst} for particles in the 
$(1/2,j_1)\oplus(j_1,1/2)$ sectors of the four-vector spinor as a function of 
$\h=\omega/m$ 
(where $\omega$ is the energy of the incident photon) up to $\h=10$. The 
long-dashed curve corresponds 
to $\widetilde{\s}(g_{(1/2^+)},\h)$ with $g_{(1/2^+)}=2$, and the short-dashed 
curve 
represents 
$\widetilde{\s}(g_{(1/2^-)},\h)$ with $g_{(1/2^-)}=-2/3$. While 
$\widetilde{\s}(g_{(1/2^+)},\h)$ is vanishing in the 
ultra relativistic limit, $\widetilde{\s}(g_{(1/2^-)},\h)$ approaches a fixed 
value, as visible from the equation (\ref{qdrtc_frm}), and also in accord  with unitarity. 
} 
\end{figure} 

 
Consequently, while in the $g_{(1/2^+)}=2$ case the scattering cross section 
for 
the genuine Dirac particle  is asymptotically vanishing, for 
$g_{(1/2^-)}=-2/3$, it approaches the fixed   $\frac{11\s_T}{27}$ value. In 
Fig.~\ref{sfig} the following quantity is plotted, 
\aeq\label{tcst} 
\widetilde{\s}(g_{(s_i)},\h)\equiv \frac{1}{\s_T}\s(g_{(s_i)},\h). 
\ceq 
{}For  $g_{(1/2^+)}=2$ one observes the usual decreasing behavior of the Dirac 
cross section 
with energy increase, 
while for $g_{(1/2^-)}=-2/3$ i.e. for 
spin-$1/2$ in   $(1/2,1)\oplus(1,1/2)$, the cross section 
$\widetilde{\s}(g_{(s_i)},\h)$ at high energy 
approaches the fixed value of $\frac{11}{27}$ as one can see in the Fig. 
\ref{sfig}.

\subsection{Calculation with the contracted equations}

The calculation of the Compton scattering with the contracted equations and associated Lagrangians is in reality equivalent to
the calculation of Compton scattering with the most general spin-$1/2$ equation (\ref{KG_gs22ref}), already studied earlier  in 
ref.~\cite{DelgadoAcosta:2010nx}, and will not be repeated here.
The result on the total cross section reported in the latter work completely and precisely coincides with our general formula for any $g_s$ given in the above equation
(\ref{qdrtc_frm}). Also the results on the Thompson limit          (\ref{Thompson}) are fully identical.
{}The revealed identity between the description of the  spin-$1/2^-$ sector in the four-vector spinor by means
of the four gauged spinorial  degrees of freedom described by the contracted equation (\ref{new_spnr}) on the one side,
with  its description by means of the gauged 16 dimensional spin-$1/2^-$ four-vector--spinors in the gauged Lagrangians in eq.~(\ref{GII}), on the other side, allows us  to conclude that
within the framework suggested,  
\begin{itemize}
\item the spin-$1/2^-$ particle in the four-vector spinor, $\psi_\mu$, indeed effectively behaves  as a truly ``quadratic'' relativistic fermion,
\item the gauging procedure conserves the number of the degrees of freedom, identified prior gauging at the free particle level by the
Poincar\'e covariant spin projectors in (\ref{comeq}), 

\item a spin-$1/2$  upon gauging, in the absence of gauged space-time symmetries and gauged covariant spin-projectors can still be identified
through the number of its gauged degrees of freedom, which in our case was shown to be four, and thereby same as the
number of the free degrees of freedom, correctly identified prior gauging  by the aforementioned Poincar\'e covariant spin projector.  
\end{itemize}

\section{Conclusions}\label{sec6} 
 
In this work we studied the $so(1,3)$ irreducible spin-$1/2$ degrees of freedom within the $(1/2^+,1/2^-,3/2^-)$ triad in the four-vector spinor, $\psi_\mu$, prior and upon gauging. 
We showed that both particles can be described by the generalized Feynman-Gell-Mann equation (\ref{KG_gs22ref}), which at the free particle level is equivalent 
to the Dirac equation. However, at the interacting level, the value of the gyromagnetic factor acquires importance and while for spin-$1/2^+$ and $g_{(1/2^+)}=2$ the quadratic equation allows for a bi-linearization towards the Dirac equation and its conjugate, for spin-$1/2^-$, characterized by $g_{(1/2^-)}=-2/3$ no linearization is possible.

The approach developed was formulated in reference to the $so(1,3)$ irreducible degrees of freedom in this space, and motivated by their interconnections by spin-up and down ladder operators, reviewed in the Appendix I. It allowed for the description of all degrees of freedom in $\psi_\mu$ on equal footing and by means of fully relativistic, representation-
and spin-specific wave equations, and associated Lagrangians, all being second order in the momenta. 
The $so(1,3)$  irreducibility turned out to be crucial for the correct
identification of the single spin-$1/2^+$ sector, $(1/2,0)\oplus (0,1/2)$, which we found identical to the Dirac particle and characterized by a gyromagnetic factor
of $g_{(1/2^+)}=2$, which allows for a bi-lineariziation. We furthermore  reproduced the precise Compton scattering results off this target known from the Dirac theory.
As a cross-check we also calculated the gyromagnetic factor for the $so(1,3)$  reducible Clebsch-Gordan combination,
$\left[{\mathcal U}^{\mathbf{SS}}(\mathbf{p},\frac{1}{2},\lambda) \right]^\a $, in
the above equation (\ref{ssdef}) and found a vanishing $g_s$ value, however we did not include this  calculation into the text for the sake of not overloading the presentation. Similarly, for  the $so(1,3)$ reducible spin-$1/2^-$ Clebsch-Gordan combinations, 
$\left[{\mathcal U}^{\mathbf{VS}}(\mathbf{p},\frac{1}{2},\lambda) \right]^\a $
in (\ref{vsdefcg}), a $g_{(1/2)}=1/3$ value has been calculated. 
 To the amount  Wigner's definition \cite{Wigner} prescribes that particles have to transform irreducibly under space-time symmetries, 
we conclude that the $so(1,3)$ reducible degrees of freedom have to be discarded as unphysical.
We recall that according to the standard definition by Wigner, fundamental particles have to 
transform according to unitary irreducible representation spaces of the Poincar\'e algebra. In field theory, 
such representations are built up as  Fourier transforms of quantum states transforming according to $so(1,3)$ irreducible finite dimensional non-unitary representation spaces of the Lorentz algebra. 
Finite-dimensional non-unitary representation spaces are of key importance in 
the description of  spin degrees of freedom and their space-time transformation properties are 
sufficient to define at the free particle level a  wave equation and corresponding Lagrangian  
in terms  of the corresponding algebra elements and invariants.\cite{DelgadoAcosta:2012yc}

Second order fermion theories have attracted recently attention through their applications within the  world line formalism in Yang-Mills theories \cite{Hostler}-\cite{Krasnov}. However, these are theories based upon two-component spinors, while our framework is based on full flashed Lorentz invariant
four-vectors and four-spinors. The information on the complete dimensionality of the  relativistic representation spaces is fully encoded by our
 respective wave equations and Lagrangians, an issue we discussed after the equations (\ref{newtree})-(\ref{newfour}).
The quadratic Lagrangians put at work in the evaluation of the Compton scattering amplitudes are such that prior gauging the spin-$1/2^-$-$3/2^-$ degrees of freedom
in the $so(1,3)$ irreducible $(1/2,1)\oplus (1,1/2)$ sector of $\psi_\mu$ have been neatly separated without auxiliary conditions at the free particle level 
by means of Poincar\'e covariant mass and spin -projectors based  on the squared Pauli-Lubanski-vector operator, a Casimir invariant of the Poincar\'e group. 
We brought an argument in support of the statement that this separation
remained respected by the gauging procedure. The argument was based on the observation that the gauged spin-$1/2^-$  Lagrangian within the full 16 dimensional space happened to provide equivalent Compton scattering description as  the 4 dimensional spinor solving the generalized Feynman--Gell-Mann equation in (\ref{KG_gs22ref}), and for $g_{(1/2^-)}=-2/3$. To the amount the equation (\ref{KG_gs22ref}) is bi-linearizable 
into  gauged Dirac equations exclusively for a gyromagnetic factor taking the special value of two units, no linearization for spin-$1/2^-$ is possible and this sector from the four-vector spinor behaves as a truly ``quadratic'',  fully relativistic fermion. As a further consequence of the observed equality we wish to note that along the line of 
ref.~\cite{Velo:1970ur}, it is easy to verify that the gauged degrees of freedom in the generalized Feynman--Gell-Mann equation (\ref{KG_gs22ref})  propagate causally within an electromagnetic environment. We expect our findings to contribute to the  understanding of the properties of the particles residing within the four-vector spinor space and shed more light on possible interferences between the spin  sectors in physical processes governed by the spin-up and down ladder operators discussed in the Appendix I.

\section*{ Appendix I: Spin-ladder operators within \texorpdfstring{$(1/2,1)\oplus (1,1/2)$}{} } 

{}From group theory it is well known  that the states spanning any irreducible 
representation space are 
related to each other by  ladder operators. Specifically, the degrees of 
freedom within  representation spaces 
of multiple spins are interconnected by  spin-up and down ladder operators. 
We here are interested in such operators within the two-spin valued space, $ (1/2,1)\oplus (1,1/2)$ 
and begin  designing them  in the rest frame, $p_0=m, \,\, {\mathbf p}= 
{\mathbf 0}$, where their form is 
 specifically simple and given by, 
\aeq 
{\mathcal K}_\pm^P(\mathbf{0})=\pm \frac{1}{\sqrt{2}}\(J_-^V\otimes J_+^S- 
J_+^V\otimes J_-^S\). 
\ceq 
Here the label $P$ on ${\mathcal K}_\pm^P$ indicates that this operator 
ladders only between eigenstates to the 
Poincar\'e projectors. Here $J_\pm^{V}$ and $J^{S}_\pm $ are in turn the 
conventional spin-up and down ladder operators in 
the respective four-vector $(1/2,1/2)$,  and the 
Dirac- spinor, $(1/2,0)\oplus (0,1/2)$,  representations, 
\aeq 
J_{\pm}^{V/S}=J_{1}^{V/S}\pm i J_{2}^{V/S}. 
\ceq 
The rest-frame four-vector-spinors, $w^{(1/2, 1)}_\pm(\mathbf{0},s_i,\lambda )$, 
with 
$s_1=1/2$, and $s_2=3/2$, 
belong to the $(1/2,1)\oplus(1,1/2)$ sector and are, 
\aec 
 w^{(1/2, 1)}_\pm\left(\mathbf{0},\frac{1}{2},\lambda \right)&=&-2\,{\mathcal 
P}_F^{(1/2, 1)}{\mathcal U}_\pm \left( {\mathbf 0}, \frac{1}{2},\l\right),\\ 
w^{(1/2, 1)}_\pm\left (\mathbf{0},\frac{3}{2},\sigma \right)&=&-{\mathcal 
P}_F^{(1/2, 1)}{\mathcal U}_\pm\left( \mathbf{0},\frac{3}{2},\sigma \right), 
\cec 
where the $(-2)$ and $(-1)$ factors have been additionally introduced for the 
sake 
of their normalization to $\pm 1$ in dependence on their parities. One can 
prove commutativity of the Poincar\'e ladder operators, 
${\mathcal K}_\pm^P(\mathbf{0})$, and the Lorentz projector ${\mathcal 
P}_F^{(1/2, 1)}$,  which can be used 
to build up Lorentz ladder operators among these states as, 
\aeq 
{\mathcal K}_\pm^L(\mathbf{0})=2 {\mathcal P}_F^{(1/2, 1)} {\mathcal 
K}_\pm^P(\mathbf{0}). 
\ceq 
The above operator annihilates  spin-$3/2$ with spin-projections $\lambda_i=\pm 3/2$, 
\aec 
{\mathcal K}_\pm^L(\mathbf{0})w^{(1/2, 1)}_\pm\left( \mathbf{0},\frac{3}
{2},\pm 
\frac{3}{2}\right) 
&=&2{\mathcal P}_F^{(1/2, 1)} {\mathcal K}^P_\pm(\mathbf{0}) 
w^{(1/2, 1)}_\pm\left(\mathbf{0},\frac{3}{2},\pm \frac{3}{2}\right)\nonumber\\ 
&=&-2{\mathcal P}_F^{(1/2, 1)} {\mathcal K}^P_\pm(\mathbf{0}) 
{\mathcal U}_\pm\left( \mathbf{0},\frac{3}{2},\pm \frac{ 3}{2}\right)=0. 
\cec 
Instead, while acting  on spin-$3/2$ with spin-projection $\l=\pm 1/2$, 
a spin-$1/2$ state is reached according to, 

\aec 
{\mathcal K}_-^L(\mathbf{0})w^{(1/2,1)}_\pm\left(\mathbf{0},\frac{3}{2},\pm 
\frac{1}{2}\right)&=&2{\mathcal 
P}_F^{(1/2, 1)} {\mathcal K}^P_-(\mathbf{0}) w^{(1/2, 
1)}_\pm\left(\mathbf{0},\frac{3} 
{2},\pm \frac{1}{2}\right),\\ 
&=&-2{\mathcal P}_F^{(1/2, 1)} {\mathcal K}^P_- 
(\mathbf{0}){\mathcal U}_\pm\left(\mathbf{0},\frac{3}{2},\pm \frac{1}
{2}\right)\\ 
&=&-2{\mathcal P}_F^{(1/2, 1)} {\mathcal U}_\pm\left(\mathbf{0},\frac{1}
{2},\pm \frac{1} 
{2}\right). 
\cec 
In effect, the spin-$3/2$ is lowered by one unit down to spin-$1/2$, 
\begin{eqnarray} 
{\mathcal K}_-^L(\mathbf{0})w^{(1/2, 1)}_\pm\left(\mathbf{0},\frac{3}{2},\pm 
\frac{1}{2}\right)&=&w^{(1/2, 1)}_\pm\left(\mathbf{0},\frac{1}{2},\pm \frac{1} 
{2}\right), 
\end{eqnarray} 
and vice verse. One can also rise the spin-$1/2$ to spin-$3/2$ as visible from, 
\aec 
{\mathcal K}_+^L(\mathbf{0})w^{(1/2, 1)}_\pm\left(\mathbf{0},\frac{1}{2},\pm 
\frac{1}{2}\right)&=&w^{(1/2, 1)}_\pm\left(\mathbf{0},\frac{3}{2},\pm \frac{1} 
{2}\right), 
\cec 
thus  confirming  ${\mathcal K}^L_\pm({\mathbf 0})$  as the rest-frame spin up and down 
ladder operators. 
The covariant spin-ladder operators in any inertial frame are then obtained 
by similarity transforming  ${\mathcal K}^L_\pm({\mathbf 0})$  by the boost 
operator 
in the representation under consideration, 
\aec 
{\mathcal K}_\pm^L(\mathbf{p})&=&B(\mathbf{p}){\mathcal 
K}_\pm^L(\mathbf{0})B^{-1}(\mathbf{p})=B(\mathbf{p}){\mathcal 
K}_\pm^L(\mathbf{0})B(\mathbf{-p}). 
\cec 
The above considerations show that one can choose  anyone of the spin degrees 
of freedom  in $(1/2,1)\oplus (1,1/2)$ as a  point of departure  for building up the 
other one. 
These facts force us to conclude that there is no kinematic reason, and as we have shown there is also no dynamical reason
for which one type of degrees of freedom in a representation space of multiple spins, has to be preferred over another type.

\section*{ Appendix II: Spin-$3/2$ within the four-vector--spinor }\label{apndx2} 
\subsection{The rigorous spin-$3/2$ third order equation,  its problems and the need for lowering its order}

The  wave equations for the spin-$3/2$ four-vector spinors from (\ref{FirstRS})-(\ref{ClebshGord}), $\[{\mathcal U}_+ ({\mathbf p},\frac{3}{2},\sigma)\]^\a$, and
$\[{\mathcal U}_- ({\mathbf p},\frac{3}{2},\sigma)\]^\a$, of positive and negative parity, respectively,  are defined through,
\begin{equation}
\left[ \Pi^{(3/2)} _\pm (p) \right]^{\alpha\beta}
\[{\mathcal U}_\pm  ({\mathbf p},\frac{3}{2},\sigma)\]_\b =\[{\mathcal U}_\pm  ({\mathbf p},\frac{3}{2},\sigma)\]^\a,
\label{32_prjRS}
\end{equation}where $\left[ \Pi^{(3/2)} _\pm (p) \right]^{\a\b} $ denotes a covariant  projector on these vector-spinors which is given by,
\begin{eqnarray} 
\left[ \Pi^{(3/2)} _\pm  (p)\right]^{\a\b} &=& 
\frac{\mp p\!\!\!/+m}{2m}\left[ \mathbb{P}^{(3/2)}(p)\right]^{\a\b}.
\label{RS-prj} 
 \end{eqnarray} 
Here  $\left[ \mathbb{P}^{(3/2)}(p)\right]^{\a\b}$ stands for the covariant spin-$3/2$ projector ~\cite{VanNieuwenhuizen:1981ae} defined in terms
of the squared Pauli-Lubanski operator from (\ref{w2rs})-(\ref{tw2rs})  as, 
\begin{eqnarray} 
[\mathbb{P}^{(3/2)}(p)]_{\a\b} 
&=& -\frac{1}{3}\left( \frac{{\mathcal W}^2(p)_{\a\b}}{p^2} +\frac{3}{4} g_{\a\b} \right),
\label{32PP}
\end{eqnarray} 
while $(\pm p\!\!\!/ +m)/(2m)$ is the covariant parity projector in the Dirac spinor space.
The  projector  $[\Pi^{(3/2)}(p)]^{\a\b}$ 
leads to a wave equation which is third order in the momenta. However, 
differential equations of an order higher than two  are both conceptually and 
technically problematic. The issue is that they lead to higher-order field theories which are plagued by  severe inconsistencies such as ghost solutions of the bad type,  kinetic terms of  wrong signs,  states of negative norms, and which violate unitarity.
Moreover, particles in such theories can suffer non local propagation. The  experimentally established fundamental  theories in contemporary physics are all based on second order Lagrangians, the Standard Model being a prominent example. Only in effective theories with over integrated fields, higher order Lagrangians may appear.
At the root of the troubles related to higher order differential equations is the so called Ostrogradskian instability \cite{Woodard} which, at the level of, say, classical mechanics, for concreteness,  predicts phase spaces of unstable orbits.{} In order to circumvent such problems,  it is desirable to lower the order of the above equation (\ref{32_prjRS})
before employing it in physical processes.
Such a lowering  can be carried out in two steps, first to second-, and then to first order.

\subsection{From third order equation  to  a quadratic }

A second order equation can be obtained by replacing in (\ref{RS-prj}) the parity operator by the mass projector, $p^2/m^2$, according to,
\begin{equation}
\frac{\mp p\!\!\!/+m}{2m}\longrightarrow \frac{p^2}{m^2},
\end{equation}
in which case (\ref{32PP}) becomes,

\begin{equation}
-\frac{p^2}{m^2}\frac{1}{3}\left( \frac{{\mathcal W}^2(p)_{\a\b}}{p^2} +\frac{3}{4}g_{\a\b} \right)
\[{\mathcal U}_\pm ({\mathbf p},\frac{3}{2},\sigma)\]^\b=\[{\mathcal U}_\pm ({\mathbf p},\frac{3}{2},\sigma)\]_\a.
\label{32_prjRS_2ndord}
\end{equation}
This path has been taken by the Ref.~\cite{Napsuciale:2006wr} and has been further  pursued  in 
 \cite{DelgadoAcosta:2009ic} and \cite{DelgadoAcosta:2012yc}.
The explicit wave equation resulting from (\ref{32_prjRS_2ndord}),  reported in \cite{Napsuciale:2006wr} in all necessary technical detail, which we here bring for completeness in its most interesting  electromagnetically  gauged form, reads
\begin{eqnarray}
{\Big( } \left(  \pi^{2}-m^{2}\right)  g_{\alpha\beta}-
ig_{(3/2)}\left(  \frac {
\sigma_{\mu\nu} \left[ \pi^\mu,\pi^\nu \right] }{4}
g_{\alpha\beta}-eF_{\a\b}~\right)
 &+&\frac{1}{3}\left(  \gamma_{\alpha}\not \pi
-4\pi_{\alpha}\right)  \pi_{\beta}\nonumber\\
&+&\frac{1}{3}
(\pi_{\alpha}\not \pi-\gamma_{\alpha}\pi^{2})\gamma_{\beta}{\Big)}\[{\mathcal U}_\pm ({\mathbf p},\frac{3}{2},\sigma)\]^\b
=0\ .\nonumber\\
\label{gauged_gc}
\end{eqnarray}
Here, $\pi_\mu$ is the minimally gauged four-momentum, $\pi_\mu=p_\mu - eA_\mu$ in (\ref{GT}).
The spin-$3/2$ gyromagnetic factor, $g_{(3/2)}$, has been fixed to  $g_{(3/2)}=2$ from the requirement on causal
propagation within an electromagnetic background, without that this equation is linearizable. In this way, it was demonstrated in \cite{Napsuciale:2006wr} that the above second order spin-$3/2$ wave equation is free from the Velo-Zwanziger problem suffered by the linear Rarita-Schwinger framework. This value has been further examined in \cite{DelgadoAcosta:2009ic} and shown to lead to finite Compton scattering differential 
cross sections in forward direction and in accord with  the desired unitarity.
It has to be emphasized that couplings of the type, $F^{\mu\nu} \sigma_{\m\nu}$ arise within the second order formalism upon minimal gauge of
the momenta in the $so(1,3)$ algebraically prescribed $\left[\partial^\mu, \partial^\nu \right]$ commutators of the type first seen in eq.~(\ref{KGG1}) and are not introduced {\it ad hoc} as non-minimal couplings as occasionally done within the Rarita-Schwinger framework \cite{Ferrara} for the sake of getting rid of the unitarity problem.  

\subsection{From quadratic equation  to a linear}

Below we show how the linear Rarita-Schwinger framework relates to (\ref{32_prjRS_2ndord}). 
Towards this goal we first recall that within  the direct-product space  $(1/2,1/2)\otimes [(1/2,0)\oplus(0,1/2)]$ of interest here, ${\mathcal 
W}_\mu(p)$ expresses 
as the direct sum of the Pauli-Lubanski vectors, $W_\mu(p)$,  and $\o_\mu (p)$, in 
the  respective  $(1/2,1/2)$- and Dirac-building blocks according to (\ref{PaLu_VS}).
With the aid of the formulas given in this equation it is straightforward to  calculate ${\mathcal W}^2_\mu(p)$ as, 
\begin{subequations}\label{old29} 
\aec 
\left[ {\mathcal W}^2(p)\right]_\alpha{} ^\beta &=& 
\o^2(p)g_\alpha{}^\beta +\left[W^2 (p)\right]_\alpha{}^\beta  + 
2(W^\mu (p))_\alpha{}^\beta w_\mu (p),\label{Pl1} 
\\ 
\o^2(p)&=&-\frac{1}{4}\sigma_{\lambda\mu} \sigma^\lambda{}_\nu p^\mu p^\nu= 
-\frac{3}{4}p^2 +\frac{1}{4}m \gamma_\eta \gamma_\lambda \left[ p^\lambda, 
p^\eta \right], \label{Pl22} 
\\ 
\left[W^2(p) \right]_\alpha{}^\beta&=& 
-2(g_{\alpha}{}^{\beta}g_{\mu\nu}  -g_{\alpha\nu}g^{\b}{}_{\mu})p^\mu p^\nu= 
-2(g_\alpha \, ^\beta p^2 -p_\alpha p^\beta),\\ 
2(W^\mu(p))_\alpha{}^\beta \o_\mu (p) &=& i\epsilon^\mu{}_\alpha{}^\beta{}_\sigma 
p^\sigma\gamma_5(p_\mu-\gamma_\mu \slashed{p}) 
=-i\epsilon^\mu{}_\alpha{}^\beta{}_{\s} p^\sigma \gamma_5\gamma_\mu 
\slashed{p}, 
\label{PL3} 
\cec 
\end{subequations} 
where we again for the sake of simplicity  suppressed the spinorial indexes. 
In effect,  the ${\mathcal W}^2(p)$ eigenvalue problem takes the following form, 
\begin{equation} 
\left( {\mathcal W}^2(p)\right)_{\alpha\beta}\[{\mathcal U}_\pm\left({\mathbf p}, 
\frac{3}{2},\sigma\right)\]^\b= 
-\frac{11}{4} p^2 \[{\mathcal U}_\pm\left({\mathbf p},\frac{3}{2},\sigma 
\right)\]_\a 
+2 \left(\omega_\mu(p) W^\mu (p)\right)_{\alpha \beta}\[{\mathcal U }_\pm \left( 
{\mathbf p}, \frac{3}{2},\sigma\right)\]^\b. 
\label{intfr1} 
\end{equation} 
In order to ensure projection on spin-$3/2$ by means of (\ref{32_prjRS_2ndord}), 
i.e. in order to guarantee  that 
the  ${\mathcal W}^2(p)$ eigenvalue, $s_2(s_2+1)$ 
in (\ref{W2eigenvalues}) equals,  $\left(-p^2 
\frac{3}{2} \left( \frac{3}{2}+1 \right)\right)$, 
it is necessary that the interference terms contributes $(-p^2)$ to the eigenvalue and satisfies, 
\begin{equation} 
2 \left(\omega_\mu (p) W^\mu (p) \right)_{\alpha \beta}\[{\mathcal U}_\pm 
\left({\mathbf p},\frac{3}{2},\sigma\right)\]^\b 
=-p^2\[{\mathcal U}_\pm \left({\mathbf p},\frac{3}{2},\sigma \right)\]_\a. 
\label{intfr1temporal} 
\end{equation} 
Substitution of (\ref{PL3}) into (\ref{intfr1}) and setting $\slashed{p}\
\left[{\mathcal U}_\pm({\mathbf p },3/2,\sigma\right]_\alpha =\mp m\left[{\mathcal U}_\pm({\mathbf p},3/2,\sigma\right]_\alpha$, 
amounts to 
\begin{equation} 
\left( 
\mp i\epsilon^\mu{}_\alpha{}^\beta{}_{\s} p^\sigma \gamma_5\gamma_\mu - m 
g_{\alpha}\, ^{\beta} +2\frac{p_\a p^\b}{m} \right) 
\[{\mathcal U}_\pm \left({\mathbf p},\frac{3}{2},\sigma  \right)\]_\b =0. 
\label{intfr2} 
\end{equation} 
The latter equation coincides, modulo the legitimate  replacement, $2p^\a p^\b/m\longrightarrow m\g^\a\g^\b$, 
allowed by the auxiliary conditions, $p\cdot {\mathcal U}_\pm \left({\mathbf p}, 3/2,\lambda  \right)=
\gamma \cdot {\mathcal U}_\pm \left({\mathbf p}, 3/2,\lambda  \right)=0$, 
with  a text-book equation that represents in a compressed form the  Rarita-Schwinger framework 
in the equations (\ref{Dirceq}), (\ref{32aux1}), and (\ref{32aux2}) and reported among others  in \cite{Lurie}.
The above considerations relate the Rarita-Schwinger framework to a restriction of the covariant mass-$m$ and spin-$3/2$ projector
based wave equation in (\ref{gauged_gc}) down to a linear operator.
Notice that as long as the spin-$3/2$ exclusively belongs to the $so(1,3)$ irreducible sector of the four-vector spinor,
the  application of the  Lorentz projector, ${\mathcal P}_F^{(1/2,1)}$ from (\ref{feq32}) is superfluous. 
The formalism systematically employed in the present work is consistent with  the second order equation (\ref{32_prjRS_2ndord}) from above.

 

\end{document}